\documentclass[final,3p,times,12pt]{elsarticle}

\usepackage{graphicx}%
\usepackage{multirow}%
\usepackage{amsmath,amssymb,amsfonts}%
\usepackage{amsthm}%
\usepackage{mathrsfs}%
\usepackage[title]{appendix}%
\usepackage{xcolor}%
\usepackage{textcomp}%
\usepackage{manyfoot}%
\usepackage{booktabs}%
\usepackage{algorithm}%
\usepackage{algorithmicx}%
\usepackage{algpseudocode}%
\usepackage{listings}%
\usepackage{tabularx}%
\usepackage{adjustbox}%
\usepackage{graphicx}
\usepackage{comment}%
\usepackage[inline]{enumitem}%
\usepackage{caption}
\usepackage{subcaption}
\usepackage[hidelinks]{hyperref}

\usepackage{fancyvrb}
\usepackage{fvextra}

\usepackage{enumitem,amssymb}
\newlist{todolist}{itemize}{2}
\setlist[todolist]{label=$\square$}
\usepackage{pifont}
\newcommand\model[1]{\texttt{#1}}

\newcommand\changed[1]{\colorlet{saved}{.}\color{black}{#1 }\color{saved}}

\newcommand\changedsec[1]{\colorlet{saved}{.}\color{black}{#1 }\color{saved}}

\newcommand\restructured[1]{\colorlet{saved}{.}\color{purple}{#1 }\color{saved}}

\makeatletter
\newcommand\footnoteref[1]{\protected@xdef\@thefnmark{\ref{#1}}\@footnotemark}
\makeatother

\theoremstyle{thmstyleone}%
\theoremstyle{thmstyletwo}%

\theoremstyle{thmstylethree}%

\journal{Information Processing \& Management}

\begin{document}

\begin{frontmatter}

\title{Large Language Models for Scholarly Ontology Generation: \\An Extensive Analysis in the Engineering Field}



\author[1]{Tanay Aggarwal\corref{cor1}} 
\ead{tanay.aggarwal@open.ac.uk}
\author[1]{Angelo Salatino} 
\ead{angelo.salatino@open.ac.uk}
\author[1,2]{Francesco Osborne} 
\ead{francesco.osborne@open.ac.uk}
\author[1]{Enrico Motta} 
\ead{enrico.motta@open.ac.uk}
\affiliation[1]{organization={Knowledge Media Institute, The Open University},
            city={Milton Keynes},
            country={UK}}
\affiliation[2]{organization={Department of Business and Law, University of Milano Bicocca},
            city={Milan},
            country={IT}}

\cortext[cor1]{Corresponding author}








\begin{abstract}
Ontologies of research topics are crucial for structuring scientific knowledge, enabling scientists to navigate vast amounts of research, and forming the backbone of intelligent systems such as search engines and recommendation systems. However, manual creation of these ontologies is expensive, slow, and often results in outdated and overly general representations. As a solution, researchers have been investigating ways to automate or semi-automate the process of generating these ontologies. One of the key challenges in this domain is accurately assessing the semantic relationships between pairs of research topics. \changedsec{This paper presents an analysis of the capabilities of large language models (LLMs) in identifying such relationships, with a specific focus on the field of engineering.
To this end, we introduce a novel benchmark based on the IEEE Thesaurus for evaluating the task of identifying three types of semantic relations between pairs of topics: \textit{broader}, \textit{narrower}, and \textit{same-as}.} Our study evaluates the performance of seventeen LLMs, which differ in scale, accessibility (open vs. proprietary), and model type (full vs. quantised), while also assessing four zero-shot reasoning strategies.
Several models with varying architectures and sizes have achieved excellent results on this task, including Mixtral-8×7B, Dolphin-Mistral-7B, and Claude 3 Sonnet, with F1-scores of 0.847, 0.920, and 0.967, respectively. Furthermore, our findings demonstrate that smaller, quantised models, when optimised through prompt engineering, can achieve strong performance while requiring very limited computational resources. 

\end{abstract}



\end{frontmatter}

\section{Introduction}\label{sec1}

\changed{Ontologies of research topics, along with other knowledge organization systems (KOS), such as taxonomies and thesauri, play a vital role in organizing and retrieving information from digital libraries~\cite{salatino_survey_2024}. Major publishers and organizations in this domain, such as Elsevier, Springer Nature, ACM, IEEE, and PubMed, employ a variety of classification systems, including the All Science Journal Classification (ASJC), SpringerNature Taxonomy, ACM Computing Classification System\footnote{ACM Computing Classification System - \url{https://dl.acm.org/ccs}}, the IEEE Theasurus, and Medical Subject Headings. These systems are crucial for enabling efficient information discovery and supporting effective scientific dissemination~\cite{Dunne2020,lipscomb2000medical,rous2012major}.} Furthermore, ontologies of research topics play a vital role in supporting a wide range of intelligent systems to navigate and interpret academic literature~\cite{osborne2013rexplore,beel2016paper,gusenbauer2020academic}. These include search engines~\cite{gusenbauer2020academic}, conversational agents~\cite{meloni2023integrating}, analytics dashboards~\cite{angioni2021aida}, and recommender systems~\cite{beel2016paper}. 

Nevertheless, generating and maintaining ontologies of research topics is both expensive and time-consuming, typically requiring the involvement of several domain experts over an extended period of time~\cite{osborne2015klink}. Moreover, the ever-increasing volume of literature, with approximately 2.5 million new scientific papers published annually~\cite{Wang_Barabasi_2021,bornmann2015}, demands continuous updates to ensure that these ontologies accurately reflect the latest research developments. 
Over the past 25 years, Artificial Intelligence (AI) researchers have developed several methods for the automated or semi-automated generation of such ontologies, with varying degrees of success~\cite{sanderson1999deriving,osborne2012mining,osborne2015klink,han2020wikicssh}. Some notable achievements include the semi-automatic pipeline for generating OpenAlex Topics~\cite{openalex2024} and the automatic development of the Computer Science Ontology~\cite{10.1162/dint_a_00055}, along with a few other initiatives. However, most of the ontologies in this domain are still manually created because current solutions still struggle to generate large-scale, detailed representations of research topics.
Consequently, this challenge remains far from being fully addressed. 



Over the past two years, Large Language Models (LLMs) have transformed the field of Natural Language Processing (NLP) by significantly improving machine comprehension of human language. In particular, some studies have demonstrated that LLMs can achieve remarkable performance even in zero-shot settings, where the model makes predictions without prior exposure to specific tasks or examples~\cite{kojima2023largelanguagemodelszeroshot}.

\changedsec{
In this paper, we explore the potential of LLMs to facilitate ontology generation by identifying semantic relationships between research topics, with a specific focus on the field of engineering.
This task is a fundamental step in the construction of scientific KOSs, and has traditionally been the most labour-intensive phase, often involving extensive deliberation among domain experts~\cite{osborne2015klink}. 
Specifically, we analyse the capability of several state-of-the-art LLMs to detect three semantic relations frequently adopted in such ontologies: \textit{broader}, \textit{narrower}, and \textit{same-as}. In particular, our experimental analysis focuses on identifying such relations between research topics defined in the IEEE Thesaurus, the most widely adopted ontology in electrical and electronic engineering, which also offers broad coverage of computer science.
To support this investigation, we introduce IEEE-Rel-1K, a gold-standard dataset derived from the IEEE Thesaurus, which describes the semantic relationships of 1,000 topic pairs.
}


We evaluated 17 off-the-shelf LLMs on this dataset, varying in scale, accessibility (open or proprietary), and model type (full or quantised). The models included 3 full open models, 10 open quantised models, and 4 proprietary models. 
For each model, we employed four distinct zero-shot prompting strategies. The first strategy involved a straightforward prompt asking the model to determine the relationship between two topics, $t_a$ and $t_b$. The second strategy extended this by applying the same method to both the original pair ($t_a$, $t_b$) and the reversed pair ($t_b$, $t_a$), with the results then being combined. The third strategy utilised a Chain-of-Thought (CoT) prompt, encouraging the model to reason through the relationship step by step. The fourth strategy was a two-way version of this CoT approach, where the process is repeated for both the original and reversed pairs before integrating the outcomes.


\changedsec{
Our findings show that several models exhibit strong zero-shot reasoning capabilities for this task. Proprietary models performed best, with Claude 3 Sonnet achieving an impressive F1 score of 0.967. Nonetheless, several open models also delivered highly competitive results, particularly when enhanced with advanced prompting techniques based on Chain-of-Thought (CoT) reasoning. Among the open models, the top performer was Dolphin-Mistral-7B (0.920 F1), a Mistral variant fine-tuned on the Dolphin dataset. The Mixtral model also demonstrated solid performance, reaching an F1 score of 0.847.

Our study produced several notable insights. First, many models exhibit strong zero-shot reasoning capabilities for this task. Second, the performance of LLMs is highly dependent on the prompting strategy employed. In particular, the two-way variant of CoT prompting consistently outperforms alternative approaches. In fact, our results show that prompting strategies can lead to substantial performance gains, with improvements exceeding 0.2 F1 points in several instances. 
Finally, when the optimal prompting strategy described in this paper is adopted, smaller models, which are more scalable and cost-efficient, can achieve performance comparable to that of the best-performing architectures.
These findings point to promising directions for the development of optimised and efficient models for ontology generation.
}




\changed{

Ultimately, our objective is to develop a detailed, comprehensive ontology of research topics across all domains. To achieve this, we plan to develop and publish field-specific ontologies using the SKOS standard\footnote{SKOS - \url{https://www.w3.org/TR/2009/REC-skos-reference-20090818/}}, subsequently mapping them together. This paper represents the first step in this ambitious direction. 
}

In summary, the contributions of this paper are as follows:
\begin{itemize}
\item We conduct a comprehensive analysis of 17 recent LLMs in identifying semantic relationships between research topics in a zero-shot setting.
\item We introduce IEEE-Rel-1K, a gold standard for research topic relation prediction, designed to facilitate the evaluation of models in this domain.
\item We explore this task under various conditions, examining different types of LLMs and evaluating the effectiveness of multiple prompting strategies.
\item We provide the complete codebase for our methodologies, including the gold standard\footnote{Gold Standard and code for experiments can be found at \url{https://github.com/ImTanay/LLM-Automatic-Ontology-Generation}}.
\end{itemize}

\color{black}

The remainder of this paper is structured as follows. Section~\ref{sec2} provides an overview of the literature. Section~\ref{background} defines the task, describes the gold standard, and presents the seventeen LLMs employed in this study. 
Section~\ref{sec:experiments}  describes the experiments and provides details on the implementation setup.
Section~\ref{sec5} presents and discusses our results. Section~\ref{erroranalysis} provides an analysis of the types of errors made by the models. \changedsec{Section~\ref{sec:limitations} discusses the key findings and the limitations.}
Finally, in Section~\ref{sec:conclusions} we conclude the paper and provide future directions.

\label{sec:limitations}

\section{Related Work}\label{sec2}

\changed{
In this section, we review the relevant literature. We begin with a discussion on knowledge organization systems (Section \ref{sec:kosr}), then discuss ontologies of research areas (Section \ref{sec:research-area-ontologies-and-taxonomies}), and finally explore state-of-the-art methodologies for generating these ontologies (Section \ref{sec:ontologygeneration}).}

\changed{
\subsection{Knowledge Organization Systems}\label{sec:kosr}

Knowledge Organisation Systems (KOSs) are frameworks for structuring information to facilitate efficient knowledge management and retrieval~\cite{zeng2008knowledge}. These systems, varying in structure and functionality (e.g., hierarchical relationships, synonym control), are commonly categorised into four primary types: term lists, taxonomies, thesauri, and ontologies.

A \textit{term list} constitutes a linear, non-hierarchical compilation of subject headings or descriptors, facilitating the organisation of document collections \cite{hedden2010taxonomies,Zaharee2013}. In literature, they are alternatively addressed as \textit{pick list}~\cite{niso2005} or as \textit{terminology}~\cite{ISO25964}. These KOSs are distinguished by the absence of explicit relational definitions between subjects.
A \textit{taxonomy} arranges classes within a hierarchical framework, with parent-child relationships \cite{rasch1987nature}. They are organised in tree structures, featuring a root node, which subsequently unfolds into sub-branches. 
A \textit{thesaurus} extends the taxonomy structure in which subjects are still organised hierarchically and are further characterised by descriptive attributes, including definitions, related terms, and synonyms~\cite{niso2005}.
Finally, an \textit{ontology} is a formal and explicit representation of domain knowledge, classifying entities based on their relevant attributes~\cite{gruber1993}. In practice, ontologies include concepts, objects, and their relations, and represent the most functionally comprehensive type of KOSs~\cite{genesereth2012}. They offer advanced capabilities, including synonym control, establishment of diverse relationship types, and definition of object properties~\cite{zeng2008knowledge}.

In the context of this work, an ontology of research areas applies this concept to the scholarly domain, systematically organising the various fields of study, their sub-disciplines, and their connections.

}

\subsection{Ontologies of Research Areas}\label{sec:research-area-ontologies-and-taxonomies}

In the literature we can find several ontologies of research areas, each differing in scope, size, depth, and curation. Notable examples include the ACM Computing Classification System, Medical Subject Headings, IEEE Thesaurus, and UNESCO Thesaurus~\cite{lipscomb2000medical,rous2012major,salatino2018computer,openalex2024}.

The Medical Subject Headings (MeSH) is a single-field ontology with over 30K concepts in the field of Medicine~\cite{lipscomb2000medical}. It is maintained by the National Library of Medicine (US), and mostly employed to organised content in the Medline digital library. It receives yearly updates with minor amendments to reflect the latest developments in the field.

The field of Computer Science is represented by the ACM Computing Classification System\footnote{The ACM Computing Classification System – \url{http://www.acm.org/publications/class-2012}}~\cite{rous2012major} and the Computer Science Ontology (CSO)~\cite{salatino2018computer}.
The former contains about 2K research topics and it is maintained by the Association for Computing Machinery (ACM) for its digital library. CSO, on the other hand, offers a more comprehensive view with over 14,000  research concepts. 
CSO was automatically generated using Klink-2~\cite{osborne2015klink}, starting from a corpus of 16 million scientific publications; and is being employed by Springer Nature to enhance the metadata quality of their computer science proceedings, including the Lecture Notes in Computer Science and Lecture Notes in Artificial Intelligence~\cite{salatino2019improving}.

In the field of Engineering, there is the IEEE Thesaurus, which contains 12K terms. Curated by the Institute of Electrical and Electronics Engineers (IEEE), it is updated annually and serves as a basis for classifying academic articles and research within the IEEE digital library.

With regard to ontologies that cover multiple fields we can find the UNESCO Thesaurus. Curated by UNESCO, it covers 4.4K subjects and it is mainly employed for indexing and searching resources within the UNESCO’s document repository. It undergoes frequent minor revisions, typically every 2-3 months.
Similarly, the ANZSRC Fields of Research (FoR) developed by the Australia and New Zealand research councils, covers around 4.4K topics across a wide range of subjects, but it receives less frequent updates ($\sim$10 years). It is employed by Digital Science to organise research content within their main datasets: Dimensions.ai and Figshare.

 OpenAlex Topics is another multi-field taxonomy with 4,800 topics, primarily used within OpenAlex, a free and open catalogue of scholarly articles; and currently at its first release~\cite{openalex2024}.

A comprehensive survey on research topic ontologies is available in~\cite{salatino_survey_2024}.

\changed{
Despite their differences, these systems consistently employ hierarchical structures, using broader/narrower relations, and manage synonyms through `same-as' or related links. Our work strategically focuses on these common relations, allowing us to build a robust framework for creating or enhancing existing ontologies. 
Furthermore, most existing systems are manually curated, and their manual development is labour-intensive, time-consuming, and costly~\cite{salatino_survey_2024}. Although automatic and semi-automatic generation methods have recently emerged, resulting in the creation of KOS such as CSO and OpenAlex Topics, these approaches have yet to fully exploit the potential of LLMs.


Finally, it is worth noting that several of these ontologies are also integrated in large-scale knowledge graphs~\cite{peng2023knowledge}, that describe the metadata of research publications (e.g., SemOpenAlex~\cite{farber2023semopenalex}, AIDA-KG~\cite{angioni2021aida}) or specific research entities and concepts (e.g., ORKG~\cite{jaradeh2019open}, 
CS-KG~\cite{dessi2022cs}, Nano-publications~\cite{kuhn2016decentralized}).

}




\subsection{Generation of Ontologies of research topics}\label{sec:ontologygeneration}

Previous decades have seen the emergence of several automated solutions for ontology creation, yet these have primarily been applied outside the scholarly domain, explaining why scholarly ontologies often remain manually curated. 
\changed{
This section is structured into three parts. First, we examine established methodologies for automatic ontology generation, with a focus on traditional natural language processing (NLP) techniques. Second, we broaden the scope to include the application of large language models (LLMs). Finally, we conclude by reviewing the currently limited applications of these techniques in the scholarly domain.
}

Early work focused on semi-automatic methods, combining human expertise with computational techniques. Maedche and Staab~\cite{maedche2001learning} introduced a method using linguistic and statistical techniques to extract concepts and relationships from text, but heavily relied on human input. Cimiano et al.~\cite{cimiano2005text2onto} advanced this with Text2Onto, integrating techniques from machine learning and natural language processing to automate extraction, though it still required substantial human validation.
Fully automatic methods emerged with OntoLearn~\cite{navigli-etal-2004-quantitative}, which used lexical resources and statistical methods to reduce human oversight. The same team~\cite{velardi2013ontolearn} later improved it with a graph-based algorithm in OntoLearn Reloaded, enhancing accuracy and robustness in complex domains. 
The integration of deep learning marked a significant advancement for the field of ontology generation. BERT~\cite{devlin2018bert} revolutionised language understanding, enhancing concept extraction and relationship identification. 
BERTopic~\cite{grootendorst2022bertopic} leveraged embeddings for coherent topic generation, influencing ontology and taxonomy extraction.
Le et al.~\cite{le2019inferring} use hyperbolic embeddings to capture the hierarchical structure of concepts. 
As recent advancements, Chen et al.~\cite{chen2020constructing} employed pre-trained models for developing hierarchical taxonomies, obtaining improved results compared to state-of-the-art solutions.  
Some efforts have also focused on the automatic or semi-automatic evolution of ontologies~\cite{canito2022systematic, osborne2018pragmatic}.

\changed{This foundation has been significantly advanced by recent research leveraging LLMs for ontology learning.  
Babaei Giglou et al.~\cite{babaei2023llms4ol} introduced LLMs4OL, demonstrating how LLMs can extract concepts, hierarchical relations, and non-taxonomic relations from text, addressing the limitations of previous statistical approaches.} 
\changed{However, their methodology suffered from critical limitations, particularly in extracting non-taxonomic relations consistently and maintaining semantic coherence across different domains, which introduces potential bias and hallucination risks.}
\changed{Building upon this foundational work, Saeedizade and Blomqvist~\cite{saeedizade2024navigating} conducted a comprehensive examination of LLMs' capabilities across the entire ontology development lifecycle. Their analysis revealed that while LLMs excel at ontology requirements gathering and conceptualisation, they still require substantial human oversight for formalisation tasks.} \changed{Critically, they show that models struggles with maintaining precise ontological consistency and their inability to autonomously validate ontological structures without human intervention.}
\changed{Kommineni et al.~\cite{kommineni2024human} proposed an innovative framework combining human expertise with LLM capabilities for knowledge graph construction. Their approach demonstrated significant efficiency improvements, reducing expert time requirements while maintaining quality.} 
\changed{Nevertheless, the study highlighted consistent challenges, including performance inconsistencies across different knowledge domain contexts and potential semantic drift during automated graph generation.}
\changed{
Zhang et al.~\cite{zhang2024ontochat} explored the integration of conversational interfaces into ontology engineering and introduced OntoChat, a framework that enables domain experts to construct ontologies through natural language dialogue.} 
\changed{While innovative, the approach revealed significant limitations, particularly in scalability for complex ontologies and the introduction of potential communication ambiguities that could compromise ontological precision.
Sun et al.~\cite{sun2024large} systematically evaluated the potential of LLMs to replace traditional taxonomies.}
\changed{Their comprehensive analysis revealed that LLMs demonstrate outstanding capabilities, but they cannot consistently replace curated taxonomies. Moreover, they found significant variations in performance across domain specialisations and challenges in maintaining long-term consistency.}

\changed{While LLMs offer clear advantages over traditional ontology generation methods, particularly in concept extraction and the identification of hierarchical relationships, producing comprehensive and detailed ontologies remains an open challenge. 
However, for this work we hypothesise that carefully designed prompting strategies, combined with LLMs' advanced language processing capabilities, can yield consistent and domain-specific modelisations.
}

\changed{With regard to generating ontologies in the scholarly domain, research endeavours have explored a variety of (semi-)automated} methods. For instance, Shen et al.~\cite{shen2018web} developed a method to discover research concepts and arrange them hierarchically. Their resulting extensive taxonomy was then implemented in Microsoft Academic Graph.
Another notable method is Klink-2~\cite{osborne2015klink}, which extracts large-scale ontologies of research topics and is currently used to generate CSO. The algorithm starts with a set of keywords and analyses their relationship based on the extent of their co-occurrence. It infers semantic relationships between keyword pairs by utilising a modified subsumption method combined with temporal information to assess hierarchical relationships, and topic similarity strategies to identify keywords referring to the same subject.
\changed{Finally, the OpenAlex team expanded the ASJC structure found in Scopus by incorporating over 4.5K new topics identified through citation clustering~\cite{openalex2024}. This involved clustering citation networks to identify coherent and thematically related groups of papers, using LLMs to label these groups, and then connecting the newly created concepts to Scopus' ASJC.}

\changed{In conclusion, LLMs are increasingly being used for ontology generation, as well as a range of research-related tasks such as identifying relevant papers~\cite{cadeddu2024comparative, khanal2020systematic}, assisting with or even automating the creation of literature reviews~\cite{bolanos2024artificial}, and enhancing academic writing and referencing~\cite{buscaldi2024citation, brody2021scite}. However, their application to research topic ontologies remains significantly underexplored. 
In this paper, we address this issue by conducting a thorough evaluation of a broad range of recent LLMs on the task of identifying relationships between scientific topics.}

\section{Background}\label{background}

This section defines the task under investigation (Section~\ref{sec:task}), describes the gold standard developed for this study (Section~\ref{sec:dataset}), and presents an overview of the LLMs used in the experiments (Section~\ref{sec:models}).


\subsection{Task definition}\label{sec:task}
This paper addresses the challenge of identifying semantic relationships between pairs of research topics, denoted in the following as $t_A$ and $t_B$. We formalise this task as a single-label, multi-class classification problem, where each topic pair is assigned  one of the following categories:  





%

\begin{itemize}
    \setlength\itemsep{0cm}
    \item \textit{broader}: $t_A$ is a parent topic of $t_B$. E.g., \textit{machine learning} is a broader area than \textit{deep neural networks};
    \item \textit{narrower}: $t_A$ is a child topic of $t_B$. E.g., \textit{distributed databases} is a  specific area within \textit{databases}. This is the inverse relationship of \textit{broader};
    \item \textit{same-as}: $t_A$ and $t_B$ can be used interchangeably to refer to same concept. E.g., \textit{haptic interface} and \textit{haptic device}. This relationship is symmetric;
    \item \textit{other}: $t_A$ and $t_B$ do not relate according to the above categories. E.g., \textit{software algorithms} and \textit{cyclones}. This category does not represent a semantic relation; rather, it serves to provide the classifier with a mechanism for labelling negative examples. 
\end{itemize}

We focused our extraction on the broader, narrower, and same-as relations, as these constitute a fundamental set for constructing a Knowledge Organization System (KOS) that effectively represents scientific disciplines~\cite{smith2020physics}. As outlined in Sections~\ref{sec:kosr} and~\ref{sec:research-area-ontologies-and-taxonomies}, these relations play a critical role in defining hierarchical structures and managing synonymy.


\subsection{Gold Standard}\label{sec:dataset}


To evaluate the performance of LLMs on this task, we created a gold standard by selecting a sample of 1,000 semantic relationships from the IEEE Thesaurus. We have named this dataset IEEE-Rel-1K (IEEE Relations). 
The IEEE thesaurus contains around 11,730 engineering, technical, and scientific terms, including IEEE-specific society terms. It is compiled from IEEE transactions, journal articles, conference papers, and standards, and reflects the vocabulary used within the engineering and scientific fields of IEEE. Specifically, we utilised the IEEE Thesaurus v1.02\footnote{A copy of version 1.02 of the IEEE Thesaurus is available at \url{https://github.com/angelosalatino/ieee-taxonomy-thesaurus-rdf/blob/main/source/ieee-thesaurus_2023.pdf}.}, which is dated back to July 2023, and is available as a PDF following the ANSI/NISO Z39.4-2021 standard~\cite{iso}.

To create IEEE-Rel-1K, we developed a Python script\footnote{Script for extracting semantic relationships in IEEE Thesaurus - \url{https://github.com/angelosalatino/ieee-taxonomy-thesaurus-rdf}.} for extracting the hierarchical structure and relationships between terms from the IEEE Thesaurus PDF file. 
\changed{Next, we randomly selected 250 relationships for \textit{broader} and \textit{narrower}. For \textit{same-as}, we used a combination of `use preferred term' and `used for' relationships as defined in the ANSI/NISO Z39.19-2005 standard~\cite{niso2005}.  While the IEEE Thesaurus offers numerous related topics, these do not consistently meet the strict definition of synonymy~\cite{niso2005}. For instance, `4G mobile communication' is related to `5G mobile communication' but while they belong to the same category of mobile networks, they represent entirely different technologies. Therefore, a team of three experts manually validated and selected only synonyms, lexical variants, and near-synonyms. This manual validation was time-consuming, limiting our dataset to 250 samples per category.} 
To complete the dataset, we randomly generated 250 pairs of topics, ensuring they were not semantically related within the IEEE Thesaurus, and labelled these as \textit{other}.

We released IEEE-Rel-1K within the GitHub repository: \url{https://github.com/ImTanay/LLM-Automatic-Ontology-Generation}.



\subsection{Large Language Models}\label{sec:models}


Our experiments included a diverse set of seventeen LLMs, all transformer-based and utilising a decoder-only architecture. However, these models exhibit variations in training parameters, quantisation levels, openness, and fine-tuning methodologies. We categorised these models into two primary groups: open (13) and proprietary (4). This distinction was necessary because we lack comprehensive information about proprietary models in terms of their internal structures and any supplementary systems that may support them. As a result, a direct comparison between open and proprietary models would not be entirely fair. 

We further divided the 13 open models into two groups: 3 full models and 10 quantised models, the latter being quantised to 8 bits. The quantised models were employed to assess the performance of more computationally efficient solutions. Since some quantised models have been fine-tuned on specific datasets, they may outperform the original models from which they were derived in certain tasks. 

Detailed descriptions of each model are provided below.

\subsubsection{Open Models}

\textbf{Mistral-7B} (shortened as \model{mistral}) is trained on Open-Web data, has 7 billion parameters, and utilises grouped-query attention and sliding window attention with a context window of 32,000 tokens~\cite{jiang2023mistral}.

\textbf{Mixtral-8×7B} (shortened as \model{mixtral}) is a sparse mixture-of-experts network incorporating 46.7 billion parameters. Only 12.9 billion parameters are used per token, whereas the remaining parameters are part of the ``expert'' layers within its sparse mixture-of-experts architecture. It has a context window of 32,000 tokens~\cite{jiang2024mixtral}.

\textbf{Llama-2-70B} (shortened as \model{llama-2}) has 70 billion parameters and a context window of 4,096 tokens. It is trained on a corpus that includes 2 trillion tokens of data from publicly available sources~\cite{touvron2023llama}.


\subsubsection{Quantised Models}

\textbf{Dolphin-2.1-Mistral-7B} (shortened as \model{dolphin-mistral}\footnote{Dolphin-2.1-Mistral-7B - \url{https://huggingface.co/TheBloke/dolphin-2.1-mistral-7B-GGUF}}) is a model with 7 billion parameters and a token context capacity of 4,096.
It is based on Mistral-7B and fine-tuned with the Dolphin\footnote{Dolphin Dataset - \url{https://huggingface.co/datasets/cognitivecomputations/dolphin}} dataset, which is an open-source implementation of Microsoft's Orca \cite{mukherjee2023orca}, with an addition of Airoboros\footnote{Airoboros - \url{https://huggingface.co/datasets/jondurbin/airoboros-2.1}} dataset.

\textbf{Dolphin-2.6-Mistral-7B-dpo-laser} (shortened as \model{dolphin-mistral-dpo}\footnote{Dolphin-2.6-Mistral-7B-dpo-laser - \url{https://huggingface.co/TheBloke/dolphin-2.6-mistral-7B-dpo-laser-GGUF}}) is based on Mistal-7B with 7 billion parameters and fine-tuned on top of Dolphin DPO using Layer Selective Rank Reduction (LASER)~\cite{sharma2023truth}. It offers a context window of 4,096 tokens.

  
\textbf{Dolphin2.1-OpenOrca-7B} (shortened as \model{dolphin-openorca}\footnote{
Dolphin2.1-OpenOrca-7B - \url{https://huggingface.co/TheBloke/Dolphin2.1-OpenOrca-7B-GGUF}}) is a model that blends Dolphin-2.1-Mistral-7B and Mistral-7B-OpenOrca models. These models were merged using the ``ties merge''~\cite{yadav2024ties} technique, keeping the same number of training parameters and token context window size, respectively 7 billion and 4,096.

\textbf{OpenChat-3.5-1210} (shortened as \model{openchat}\footnote{OpenChat-3.5-1210 - \url{https://huggingface.co/TheBloke/openchat-3.5-1210-GGUF}}) is a model fine-tuned on top of \textit{Mistral-7B-v0.1}. It possesses similar properties and consists of 7 billion parameters with a context window size of 8,192 tokens~\cite{wang2023openchat}.

\textbf{OpenChat-3.5-0106-Gemma} (shortened as \model{Openchat-gemma}\footnote{OpenChat-3.5-0106-Gemma - \url{https://huggingface.co/gguf/openchat-3.5-0106-gemma-GGUF}}) is a model trained on \textit{openchat-3.5-0106}\footnote{openchat-3.5-0106 - \url{https://huggingface.co/openchat/openchat-3.5-0106}} 
data using Conditioned Reinforcement Learning Fine-Tuning (C-RLFT) framework \cite{de2020fine, wang2023openchat}. This model shares the same properties of the \textit{openchat-3.5-0106} model, which was fine-tuned\footnote{\url{https://huggingface.co/openchat/openchat-3.5-1210/discussions/4\#658288f1168803bdee13d6b3}} on top of Mistral-7B. It consists of 7 billion parameters with a context window of 8,192 tokens.

\textbf{SOLAR-10.7B-Instruct-v1.0} (shortened as \model{solar}\footnote{SOLAR-10.7B-Instruct-v1.0 - \url{https://huggingface.co/TheBloke/SOLAR-10.7B-Instruct-v1.0-GGUF}}) a 10.7 billion parameters model with a Depth Up-Scaling (DUS) architecture that includes architectural modifications and continued pretraining \cite{kim2023solar}. It is trained on the OpenOrca\footnote{OpenOrca dataset - \url{https://huggingface.co/datasets/Open-Orca/OpenOrca}} dataset.

\textbf{Mistral-7B-OpenOrca} (shortened as \model{mistral-openorca}\footnote{Mistral-7B-OpenOrca - \url{https://huggingface.co/Open-Orca/Mistral-7B-OpenOrca}}) is based on \textit{Mistral-7B-Instruct-v0.1} and has been fine-tuned using the OpenOrca dataset \cite{mukherjee2023orca}. This dataset is a comprehensive collection of augmented Fine-tuned LAnguage Net (FLAN) data aligned with the distributions outlined in the Orca paper \cite{lian2023mistralorca1, longpre2023flan}. The model has 7 billion parameters and a context window of 4,096.

\textbf{Eurus-7b-sft} (shortened as \model{eurus}\footnote{Eurus-7b-sft - \url{https://huggingface.co/mradermacher/Eurus-7b-sft-GGUF}}) is fine-tuned on top of \textit{Mistral-7B-Instruct-v0.2} using all correct actions in UltraInteract, mixing a small proportion of UltraChat, ShareGPT, and OpenOrca examples, and posses 7 billion parameters with a context window size of 4,096 \cite{yuan2024advancing}.


\textbf{Llama-3-8B-Instruct} (shortened as \model{llama-3}\footnote{Llama-3 - \url{https://huggingface.co/lmstudio-community/Meta-Llama-3-8B-Instruct-GGUF}}) has 8 billion parameters and a context length of almost 8,192 tokens \cite{llama3modelcard}. It utilises supervised fine-tuning (SFT) and reinforcement learning with human feedback (RLHF).

\textbf{Orca-2-13B} (shortened as \model{orca-2}\footnote{Orca-2-13B - \url{https://huggingface.co/TheBloke/Orca-2-13B-GGUF}}) is a model with 13 billion parameters and a 2,048-token context window. It was trained on a mix of publicly available and curated datasets to enhance its reasoning and comprehension abilities. The fine-tuning process was based on datasets like Super-NaturalInstructions and FLAN to ensure strong performance across NLP tasks \cite{mitra2023orca}.

\subsubsection{Proprietary Models}
We considered four proprietary models. Due to undisclosed details, we can only provide limited information. For example, their number of parameters is not officially disclosed.

\textbf{GPT-3.5-Turbo-Instruct} (shortened as \model{gpt-3.5}) is trained using RLHF, following similar methods as \textit{InstructGPT}, but with some differences in the data collection setup. It has a context window size of 4,096 tokens and was trained with data up to September 2021~\cite{ye2023comprehensive}.
  
\textbf{GPT-4-Turbo} (shortened as \model{gpt-4}) has been pre-trained to predict the next token in a document. It was trained using publicly available data and data licensed from third-party providers up to December 2023. The model was then fine-tuned using RLHF. It has a context window of 128,000 tokens~\cite{openai2023gpt4}.

\textbf{Claude 3 Haiku} (shortened as \model{haiku}) is a model with a context size of 200,000 tokens, offering a maximum output of 4,096 tokens with training data up to August 2023~\cite{anthropic}.

\textbf{Claude 3 Sonnet} (shortened as \model{sonnet}) offers a context size of 200,000 tokens. It has the capability to generate a maximum output of 4,096 tokens and has been trained on data up to August 2023. While \model{haiku} is optimised for speed and cost-effectiveness in applications requiring immediate responses, \model{sonnet} features a larger model architecture giving it advantage in handling complex tasks that require in-depth understanding of language~\cite{anthropic}.






\section{Experiments}\label{sec:experiments}


\begin{figure*}[t!]
    \centering
    \begin{subfigure}[t]{0.5\textwidth}
        \centering
        \includegraphics[width=\linewidth]{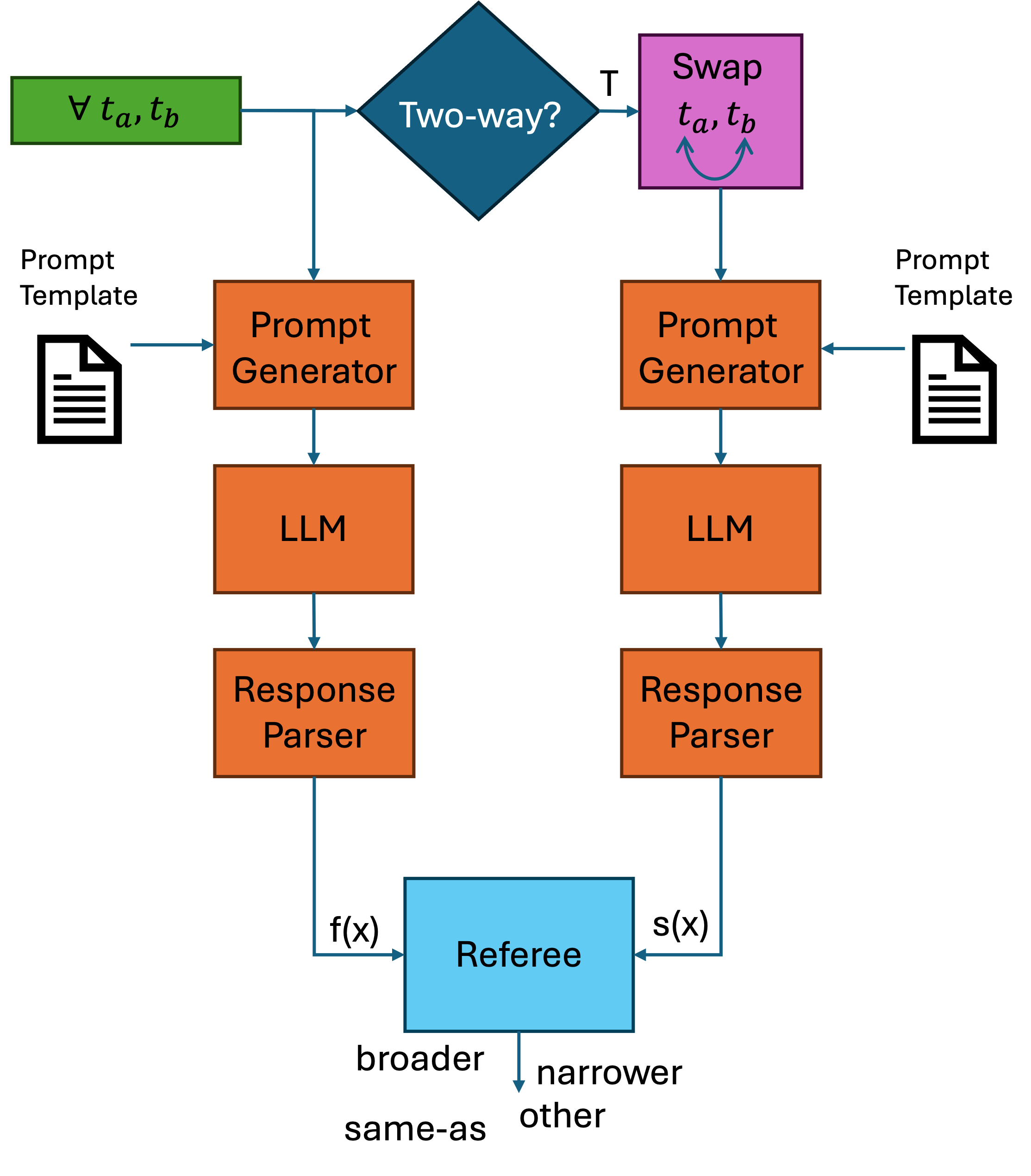}
        \caption{Standard prompting}
        \label{fig:architecturesta}
    \end{subfigure}%
    ~ 
    \begin{subfigure}[t]{0.5\textwidth}
        \centering
        \includegraphics[width=\linewidth]{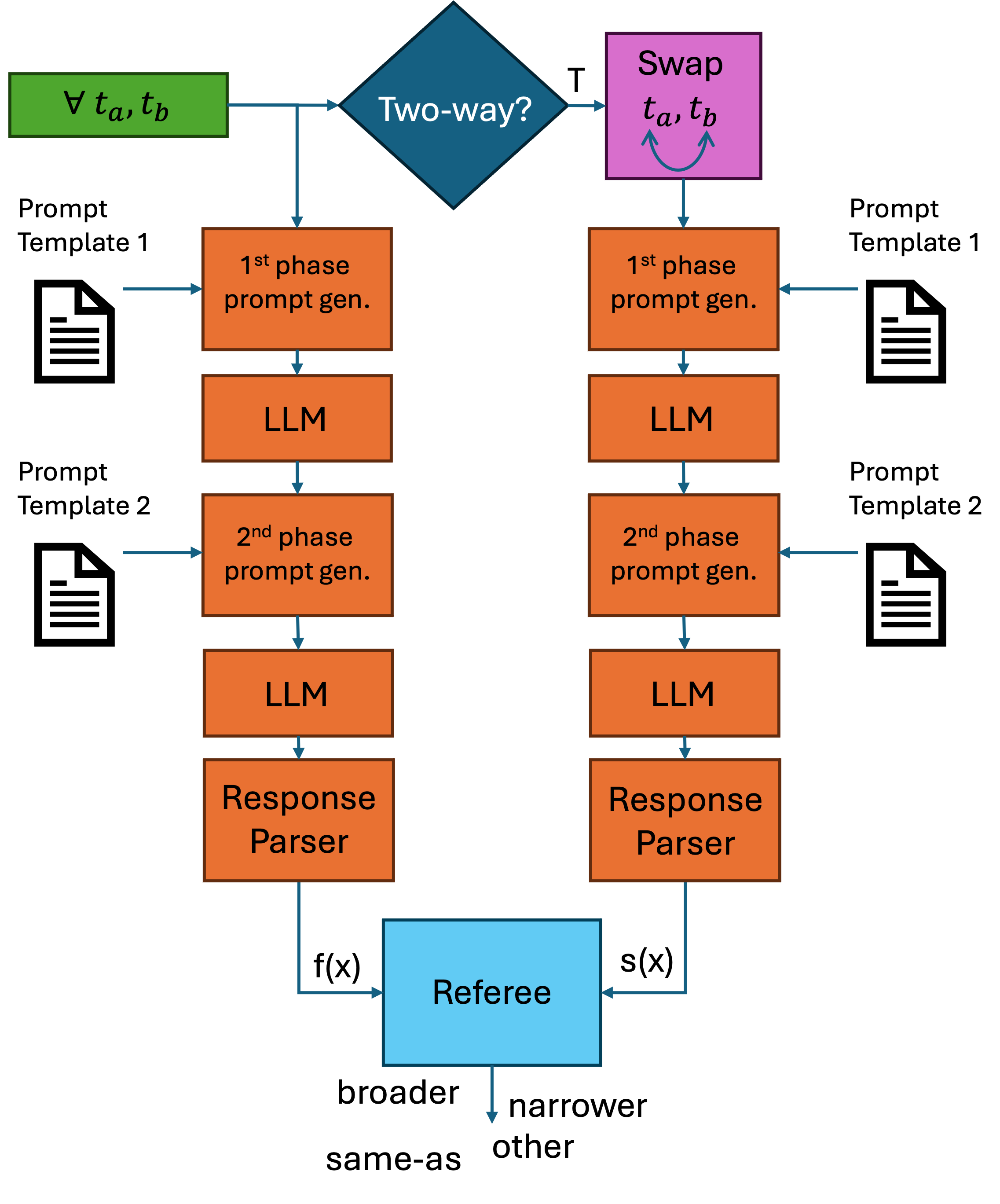}
        \caption{Chain-of-thought prompting}
        \label{fig:architecturecot}
    \end{subfigure}
    \caption{Flowcharts of our experiments. One the left there is the flowchart of our standard prompting experiments. On the right is the flowchart of our chain-of-thought prompting experiments. All boxes except the orange are similar across experiments.}
    \label{fig:architecture}
\end{figure*}

We evaluated the 17 models described in Section~\ref{sec:models} against IEEE-Rel-1K, the gold standard defined in Section~\ref{sec:dataset}, employing four different experiments that varied based on the prompting strategy utilised.

\changed{
We investigated two primary prompting paradigms: a standard prompting approach based on a simple prompt that directly asks the model to perform the classification and Chain-of-Thought (CoT) prompting~\cite{wei2022chain}, which encourages the model to reason through a predefined sequence of steps before providing an answer.  
For each of these two paradigms, we employed two additional strategies, resulting in four distinct experimental settings. The first, labelled  \textit{one-way}, involves the model inferring the relationship between $t_A$ and $t_B$ directly. The second, referred to as \textit{two-way}, builds upon the first by having the model initially infer the relationship between $t_A$ and $t_B$, then infer the relationship between $t_B$ and $t_A$, and finally combine the results using heuristic rules. 

Figure~\ref{fig:architecture} illustrates the flowcharts of these experiments.}

\changed{Based on these strategies, we can characterise our four experiments as follows:}
\begin{enumerate}
    \setlength\itemsep{0cm}
    \item standard prompting, one-way strategy (left branch of Fig.~\ref{fig:architecturesta});
    \item standard prompting, two-way strategy (both branches of Fig.~\ref{fig:architecturesta});
    \item chain-of-thought prompting, one-way strategy (left branch of Fig.~\ref{fig:architecturecot});
    \item chain-of-thought prompting, two-way strategy (both branches of Fig.~\ref{fig:architecturecot}). 
\end{enumerate}

\changed{
 In all experiments, the performance of the model was assessed using standard precision, recall and F1.

The standard prompting, one-way strategy serves as a baseline for classification tasks. We introduced CoT prompting due to its demonstrated effectiveness in complex reasoning tasks~\cite{wei2022chain, kojima2022large}. Indeed, as discussed in Section~\ref{sec:results}, CoT prompting led to improved performance for most models.
We investigated the two-way validation strategy to identify and rectify potential inconsistencies. It draws inspiration from prior research showing that approaches to improve logical consistency can enhance the overall performance of LLMs~\cite{liu2024aligning}. This approach leverages the inverse relationship between \textit{broader} and \textit{narrower}, as well as the symmetric nature of the \textit{same-as} and \textit{other} relationships~\cite{niso2005}. Conflicting answers suggest inconsistencies that may be addressed using heuristic methods.

In the following subsections, we provide a detailed description of each strategy, followed by a discussion of the experimental setup and implementation details.
}

\subsection{Standard prompting}

In the experiments using standard prompting (Fig.~\ref{fig:architecturesta}), we generated a prompt for each pair of research topics from the gold standard using a predefined template. The prompt template (detailed in ~\ref{app:promtpstandard}) describes the task and clearly defines the four categories, providing one generic example for each. Additionally, the template specifies that the answer should be provided in a numeric code format to facilitate parsing and ensure unambiguous responses. The resulting prompts were then submitted to the LLM to obtain an answer for each pair of topics. To ensure a fair comparison, all models received the same prompt.


\subsection{Chain-of-thought prompting}

CoT prompting experiments operated similarly to the previous solution but involved a two-phase interaction with the LLM, as illustrated in Fig.~\ref{fig:architecturecot}. 
In the first phase, the model is asked to provide a definition of both topics, formulate a sentence that incorporates both of them, and finally discuss their potential semantic relationships. The resulting response is then fed back into the model in the second phase, with the sole task of determining the type of semantic relationship. The two templates used to produce the prompts for the CoT experiments are available in~\ref{app:promtpcot}.




\subsection{One-way vs. two-way strategies}


The experiments for both standard prompting and CoT prompting are conducted in two different configurations. 
The one-way strategy simply asks for the relationship between $t_A$ and $t_B$. 
The two-way strategy involves asking the LLM to identify the relationship between topics $t_A$ and $t_B$, as well as the relationship between topics $t_B$ and $t_A$, where the order of the topics is reversed. This approach can naturally produce inconsistent responses, which may indicate potential issues. 

In order to detect and mitigate logical inconsistencies, we designed a set of empirical rules (cyan boxes as referees in both Fig.~\ref{fig:architecturesta} and \ref{fig:architecturecot}). The rules were designed to prioritise the development of a hierarchical taxonomy.

Considering f(X) and s(X) as the relationship types respectively obtained via the \underline{f}irst and \underline{s}econd branches of the two-way strategy; and len($t_A$) as the length of the topic's surface form, the rules are established as follows:
\begin{enumerate}
    \setlength\itemsep{0cm}
    \item broader :- f(broader) $\land$ s(narrower)
    \item narrower :- f(narrower) $\land$ s(broader)
    \item broader :- ((f(narrower) $\land$ s(narrower)) $\lor$ (f(broader) $\land$ s(broader))) $\land$ len($t_A$) $\leq$ len($t_B$)
    \item narrower :- ((f(narrower) $\land$ s(narrower)) $\lor$ (f(broader) $\land$ s(broader))) $\land$ len($t_A$) $>$ len($t_B$)
    \item same-as :- f(same-as) $\land$ s(same-as)
    \item broader :- (f(broader) $\land$ s(other)) $\lor$ (f(other) $\land$ s(narrower))
    \item narrower :- (f(narrower) $\land$ s(other)) $\lor$ (f(other) $\land$ s(broader))
    \item :- f(X) 
\end{enumerate}

\bigskip

\subsection{Experimental setup}



To interact with the various LLMs we employed two services, e.g., Amazon Bedrock\footnote{Amazon Bedrock - \label{bedrock}\url{https://aws.amazon.com/bedrock}}, and OpenAI API\footnote{OpenAI API - \url{https://platform.openai.com/docs/overview}}; as well as the KoboldAI\footnote{Kobold API - \label{kolbold}\url{https://github.com/KoboldAI/KoboldAI-Client}} library for running LLMs locally.


Specifically, through \textbf{Amazon Bedrock}, we interacted with MistralAI family LLMs (e.g., Mistral-7B and Mixtral-8$\times$7B), Anthropic models (e.g., Cluade 3 Haiku, and Claude 3 Sonnet), and Llama-2-70B by Meta. This facility provides a pre-structured generic wrapper, which encapsulates the interaction with the different LLMs offered by Amazon Bedrock.

With the \textbf{OpenAI} API we interacted with GPT models, such as GPT3.5-Turbo-Instruct and GPT4-Turbo.

Finally, we employed \textbf{KoboldAI} to interface with Dolphin-2.1-Mistral-7B, Mistral-7B-OpenOrca, Dolphin-2.6-Mistral-7B-dpo-laser, Dolphin-2.1-OpenOrca-7B,  SOLAR-10.7B-Instruct-v1.0,  OpenChat-3.5-1210, OpenChat-3.5-0106-Gemma, EURUS-7B-SFT. KoldbolAI is an open-source tool built on llama.cpp and provides model access through an API endpoint. We run it on a Google Colaboratory instance equipped with Nvidia's V100 and L4 GPUs.

Within the GitHub repository, we report the various parameters we set when interacting with the models across the three services.

\section{Results and Discussion}\label{sec5}\label{sec:results}

This section presents the results of the complete set of experiments described in Section~\ref{sec:experiments}. 
First we will discuss the standard prompting experiments (see Fig.~\ref{fig:architecturesta} for flowchart) and then we will report the results of the CoT prompting experiments (see Fig.~\ref{fig:architecturecot}).
Further, we will compare the results across the various strategies.


\subsection{Standard prompting, one-way strategy}\label{sec:s1}


Table~\ref{uni_sta} reports the results of the experiments with standard prompting when applying one-way strategy.
Among the full open models (first three rows), \model{llama-2} and \model{mixtral} demonstrate good performance in terms of average F1 with 0.669 and 0.779, respectively.
\changed{In particular, \model{mixtral} shows a high value of precision with the broader (0.819) and same-as (1), whereas \model{llama-2} obtains high values of precision with narrower (1) and other (0.877). }Complementarily, \model{mixtral} reaches high values of recall for narrower (0.888) and other (0.940), whereas \model{llama-2} for broader (0.872) and same-as (0.992). In this case, \model{mistral} consistently performs low, except for the recall of the relationship other (0.992).
\changed{The observed performance differences may be partially attributed to variations in the pre-training datasets. Indeed, \model{mixtral}~\cite{jiang2023mistral} was trained on a 
more diverse corpus that included scientific and technical documents~\cite{jiang2024mixtral}. This broader exposure may have enhanced its ability to model the relationships between scientific domains, resulting in superior performance in identifying hierarchical structures.
In contrast, \model{llama-2}~\cite{touvron2023llama} was pre-trained primarily on web data.  
}


Among quantised models, \model{dolphin-openorca} and \model{openchat} stand out with the highest average F1-score of 0.724 and 0.705, respectively. While both consistently excel in precision, they face some challenges in recall for narrower and same-as relations. Other quantised models, although falling below 0.7 in average F1-score, still demonstrate remarkable performance in specific relationships. For instance, \model{mistral-openorca} excels in precision for the same-as (1) and other (0.992) relations and recall for narrower (0.996), despite lower performance in all other metrics. 
\changed{The varied performance among models with the same base architecture but different fine-tuning approaches highlights the critical role of instruction-tuning datasets~\cite{zhang2023instruction}. \model{dolphin-openorca} and \model{openchat} were both fine-tuned on datasets that emphasise following detailed instructions~\cite{mukherjee2023orca, wang2023openchat}, which may make them more suitable for our specific task. }


When evaluating the proprietary models (as shown in the last four rows of Table~\ref{uni_sta}), \model{sonnet} stands out, achieving an outstanding average F1-score of 0.967, with a well-balanced precision and recall across all relationship types. \model{gpt-4} also demonstrates strong performance (0.948 F1). These two models exhibit high precision for same-as and other relations, and high recall for broader relations. In contrast, the two other models do not perform as well. \model{Haiku}, with an F1-score of 0.712, is surpassed by both \model{mixtral} and \model{dolphin-openorca}. Surprisingly, \model{gpt-3.5} also shows lower-than-expected results.
\changed{The exceptional performance of \model{sonnet} and \model{gpt-4} is most likely attributable to both their size and their pre-training on datasets that include extensive scientific and technical corpora~\cite{anthropic, openai2023gpt4}. While the specific composition of their training data remains proprietary, these models have demonstrated superior capabilities on several tasks that require a nuanced understanding of scientific domains~\cite{bubeck2023sparks}.
Naturally, both model size and recency play a significant role, as evidenced by the superior performance of the most recent and largest proprietary models (\model{gpt-4}, \model{sonnet}) compared to their earlier and smaller counterparts (\model{gpt-3.5}, \model{haiku}).}

\changed{Overall, we observe a clear performance gap between full open models and proprietary models. For example, \model{sonnet} achieves an average F1-Score of 0.967, substantially outperforming \model{mixtral} which has an average F1-Score of 0.779. 
}

\changed{While none of the models explicitly claim to have been trained on the IEEE digital library specifically, the superior performance of proprietary models suggests they may have encountered scientific publications during pre-training~\cite{zhao2023survey}. Kandpal et al.~\cite{kandpal2022deduplicating} demonstrated that LLMs can effectively memorise portions of their training data, particularly for well-structured technical content. }

\begin{table}[!h]
\centering
\scriptsize
\caption{Precision, recall, and F1-score for standard prompting, one-way strategy. BR are the performance associated to \textit{broader} relationships, NA to \textit{narrower}, SA to \textit{same-as}, OT to \textit{others}, and AVG are average performance considering the four types of relationships. In \textbf{bold} are the best performing models for a given class.\label{uni_sta}}
\resizebox{\columnwidth}{!}{%
\begin{tabular}{l|rrrrr|rrrrr|rrrrr}
\toprule
\multicolumn{1}{c|}{\multirow{2}{*}{\textbf{MODEL}}} & \multicolumn{5}{c|}{\textbf{PRECISION}} & \multicolumn{5}{c|}{\textbf{RECALL}} & \multicolumn{5}{c}{\textbf{F1-SCORE}} \\
\multicolumn{1}{c|}{} & \multicolumn{1}{l}{AVG} & \multicolumn{1}{l}{BR} & \multicolumn{1}{l}{NR} & \multicolumn{1}{l}{SA} & \multicolumn{1}{l|}{OT} & \multicolumn{1}{l}{AVG} & \multicolumn{1}{l}{BR} & \multicolumn{1}{l}{NR} & \multicolumn{1}{l}{SA} & \multicolumn{1}{l|}{OT} & \multicolumn{1}{l}{AVG} & \multicolumn{1}{l}{BR} & \multicolumn{1}{l}{NR} & \multicolumn{1}{l}{SA} & \multicolumn{1}{l}{OT} \\
\midrule
mistral & 0.401 & 0.667 & 0.000 & 0.511 & 0.428 & 0.462 & 0.008 & 0.000 & 0.848 & \textbf{0.992} & 0.313 & 0.016 & 0.000 & 0.638 & 0.598 \\
mixtral & \textbf{0.817} & \textbf{0.819} & 0.634 & \textbf{1.000} & 0.816 & \textbf{0.781} & 0.688 & \textbf{0.888} & 0.608 & 0.940 & \textbf{0.779} & 0.748 & \textbf{0.740} & \textbf{0.756} & 0.874 \\
llama-2 & 0.809 & 0.796 & \textbf{1.000} & 0.565 & \textbf{0.877} & 0.722 & \textbf{0.872} & 0.140 & \textbf{0.992} & 0.884 & 0.669 & \textbf{0.832} & 0.246 & 0.720 & \textbf{0.880} \\ \hline
dolphin-mistral & 0.676 & 0.613 & 0.736 & 0.901 & 0.455 & 0.599 & 0.652 & 0.412 & 0.508 & 0.824 & 0.599 & 0.632 & 0.528 & 0.650 & 0.586 \\
dolphin-mistral-dpo & 0.734 & 0.613 & 0.868 & 0.960 & 0.497 & 0.603 & \textbf{0.988} & 0.236 & 0.284 & 0.904 & 0.552 & 0.757 & 0.371 & 0.438 & 0.641 \\
dolphin-openorca & 0.780 & 0.612 & 0.810 & 0.960 & 0.737 & \textbf{0.733} & 0.952 & 0.496 & 0.576 & 0.908 & \textbf{0.724} & 0.745 & 0.615 & 0.720 & \textbf{0.814} \\
openchat & \textbf{0.826} & \textbf{0.930} & \textbf{0.991} & 0.902 & 0.481 & 0.701 & 0.740 & 0.416 & 0.664 & \textbf{0.984} & 0.705 & \textbf{0.824} & 0.586 & \textbf{0.765} & 0.646 \\
openchat-gemma & 0.679 & 0.396 & 0.471 & \textbf{1.000} & 0.849 & 0.523 & 0.968 & 0.296 & 0.268 & 0.560 & 0.506 & 0.562 & 0.364 & 0.423 & 0.675 \\
solar & 0.723 & 0.642 & 0.877 & 0.850 & 0.523 & 0.657 & 0.916 & 0.544 & \textbf{0.784} & 0.384 & 0.646 & 0.755 & \textbf{0.672} & 0.529 & 0.627 \\
mistral-openorca & 0.815 & 0.866 & 0.401 & \textbf{1.000} & \textbf{0.992} & 0.605 & 0.568 & \textbf{0.996} & 0.384 & 0.472 & 0.613 & 0.686 & 0.572 & 0.555 & 0.640 \\
eurus & 0.596 & 0.410 & 0.638 & 0.424 & 0.911 & 0.493 & 0.984 & 0.204 & 0.332 & 0.452 & 0.466 & 0.579 & 0.309 & 0.372 & 0.604 \\
llama-3 & 0.715 & 0.560 & 0.446 & \textbf{1.000} & 0.853 & 0.582 & 0.732 & 0.808 & 0.256 & 0.532 & 0.568 & 0.634 & 0.575 & 0.408 & 0.655 \\
orca-2 & 0.500 & 0.342 & 0.382 & 0.486 & 0.788 & 0.456 & 0.944 & 0.052 & 0.068 & 0.760 & 0.372 & 0.502 & 0.091 & 0.119 & 0.774 \\ \hline
gpt-3.5 & 0.668 & 0.267 & 0.407 & \textbf{1.000} & \textbf{1.000} & 0.295 & 0.976 & 0.096 & 0.032 & 0.076 & 0.195 & 0.419 & 0.155 & 0.062 & 0.141 \\
gpt-4 & 0.952 & 0.919 & 0.934 & \textbf{1.000} & 0.954 & 0.949 & \textbf{1.000} & 0.968 & 0.832 & \textbf{0.996} & 0.948 & 0.958 & 0.951 & 0.908 & \textbf{0.975} \\
haiku & 0.820 & 0.496 & 0.870 & \textbf{1.000} & 0.912 & 0.710 & 0.980 & 0.564 & 0.468 & 0.828 & 0.712 & 0.659 & 0.684 & 0.638 & 0.868 \\
sonnet & \textbf{0.969} & \textbf{0.929} & \textbf{0.950} & 0.996 & \textbf{1.000} & \textbf{0.967} & 0.996 & \textbf{0.996} & \textbf{0.932} & 0.944 & \textbf{0.967} & \textbf{0.961} & \textbf{0.973} & \textbf{0.963} & 0.971 \\ \bottomrule
\end{tabular}%
}
\end{table}

\subsection{Standard prompting, two-way strategy}

Table~\ref{bid_sta} presents the performance of the models when using standard prompting with the two-way strategy.

Among the full open models, \model{mixtral} achieves an average F1-score of 0.847, significantly outperforming both \model{llama-2} and \model{mistral}. 
It is able to identify the broader, narrower, and same-as with F1-score around 0.82, but shows exceptional capabilities (0.925) in determining the relationship other.
\changed{
The superior performance of \model{mixtral} in the two-way classification strategy likely arises from its mixture-of-experts architecture, which can dynamically activate different subsets of experts depending on the input~\cite{jiang2024mixtral}. Therefore, \model{mixtral} may be able to selectively engage experts specialized in reasoning along specific hierarchical directions~\cite{zadouri2023pushing}.
}

Regarding the quantised models, there are three models (e.g., \model{dolphin-mistral-dpo}, \model{dolphin-openorca}, and \model{openchat}) that obtain an average F1-score above 0.78.
\changed{The strong performance of these quantised models when employing the two-way strategy likely reflects recent advancements in their fine-tuning methodologies.
Specifically, \model{dolphin-mistral-dpo} benefits from Direct Preference Optimization (DPO), a technique that improves the model's capacity to follow complex instructions~\cite{rafailov2023direct}, while \model{dolphin-openorca} leverages demonstration-based learning from more capable teacher models~\cite{mukherjee2023orca}.}


Among the proprietary models, \model{sonnet} obtained comparable results to previous experiments (0.965 F1), where the performance of \model{gpt-3.5}, \model{gpt-4}, and \model{haiku} obtained a slight improvement. 


In this case, the full open models and their quantised counterparts exhibit similar performance levels (e.g., 0.847 for \model{mixtral} and 0.853 for \model{dolphin-openorca}). Moreover, the two-way strategy has effectively reduced the performance gap between these models and their larger, proprietary counterparts, narrowing the difference to just 0.113 F1 points. 
\changed{This enhancement suggests that the two-way strategy is effective in mitigating the performance gap between small and large models. This finding aligns with prior research demonstrating that similar self-verification mechanisms can improve model accuracy in information extraction tasks~\cite{gero2023self}. Indeed, the significant performance gains observed in open models indicate that they may be capable of accessing relevant knowledge when prompted within an appropriate verification framework.}

\begin{table}[!h]
\centering
\caption{Precision, recall, and F1-score for standard prompting, two-way strategy. BR are the performance associated to \textit{broader} relationships, NA to \textit{narrower}, SA to \textit{same-as}, OT to \textit{others}, and AVG are average performance considering the four types of relationships. In \textbf{bold} are the best performing models for a given class.\label{bid_sta}}
\resizebox{\columnwidth}{!}{%
\begin{tabular}{l|rrrrr|rrrrr|rrrrr}
\toprule
\multicolumn{1}{c|}{\multirow{2}{*}{\textbf{MODEL}}} & \multicolumn{5}{c|}{\textbf{PRECISION}} & \multicolumn{5}{c|}{\textbf{RECALL}} & \multicolumn{5}{c}{\textbf{F1-SCORE}} \\
\multicolumn{1}{c|}{} & \multicolumn{1}{l}{AVG} & \multicolumn{1}{l}{BR} & \multicolumn{1}{l}{NR} & \multicolumn{1}{l}{SA} & \multicolumn{1}{l|}{OT} & \multicolumn{1}{l}{AVG} & \multicolumn{1}{l}{BR} & \multicolumn{1}{l}{NR} & \multicolumn{1}{l}{SA} & \multicolumn{1}{l|}{OT} & \multicolumn{1}{l}{AVG} & \multicolumn{1}{l}{BR} & \multicolumn{1}{l}{NR} & \multicolumn{1}{l}{SA} & \multicolumn{1}{l}{OT} \\
\midrule
mistral & 0.239 & 0.000 & 0.000 & 0.451 & 0.503 & 0.475 & 0.000 & 0.000 & 0.908 & \textbf{0.992} & 0.318 & 0.000 & 0.000 & 0.603 & 0.668 \\
mixtral & \textbf{0.865} & 0.751 & 0.773 & \textbf{1.000} & 0.935 & \textbf{0.846} & \textbf{0.904} & \textbf{0.856} & 0.708 & 0.916 & \textbf{0.847} & \textbf{0.820} & \textbf{0.812} & \textbf{0.829} & \textbf{0.925} \\
llama-2 & 0.759 & \textbf{0.832} & \textbf{0.819} & 0.426 & \textbf{0.959} & 0.621 & 0.336 & 0.308 & \textbf{0.996} & 0.844 & 0.605 & 0.479 & 0.448 & 0.596 & 0.898 \\ \hline
dolphin-mistral & 0.718 & 0.686 & 0.673 & 0.884 & 0.629 & 0.704 & 0.708 & 0.732 & 0.704 & 0.672 & 0.706 & 0.697 & 0.701 & 0.764 & 0.664 \\
dolphin-mistral-dpo & 0.834 & 0.759 & 0.764 & 0.968 & 0.844 & 0.808 & \textbf{0.956} & \textbf{0.932} & 0.476 & 0.868 & 0.795 & \textbf{0.846} & \textbf{0.840} & 0.638 & 0.856 \\
dolphin-openorca & \textbf{0.862} & 0.799 & 0.774 & 0.928 & 0.947 & \textbf{0.852} & 0.888 & 0.892 & \textbf{0.776} & 0.852 & \textbf{0.853} & 0.841 & 0.829 & \textbf{0.845} & \textbf{0.897} \\
openchat & 0.838 & \textbf{0.979} & \textbf{0.964} & 0.823 & 0.586 & 0.777 & 0.736 & 0.648 & 0.744 & \textbf{0.980} & 0.783 & 0.840 & 0.775 & 0.782 & 0.734 \\
openchat-gemma & 0.734 & 0.484 & 0.512 & \textbf{1.000} & 0.939 & 0.590 & 0.804 & 0.792 & 0.332 & 0.432 & 0.579 & 0.605 & 0.622 & 0.499 & 0.592 \\
solar & 0.723 & 0.731 & 0.753 & 0.784 & 0.623 & 0.715 & 0.836 & 0.816 & 0.508 & 0.700 & 0.710 & 0.780 & 0.783 & 0.617 & 0.659 \\
mistral-openorca & 0.781 & 0.542 & 0.584 & \textbf{1.000} & \textbf{1.000} & 0.650 & 0.888 & 0.908 & 0.492 & 0.312 & 0.630 & 0.673 & 0.711 & 0.660 & 0.476 \\
eurus & 0.587 & 0.480 & 0.495 & 0.374 & \textbf{1.000} & 0.489 & 0.584 & 0.576 & 0.492 & 0.304 & 0.488 & 0.527 & 0.532 & 0.425 & 0.466 \\
llama-3 & 0.746 & 0.476 & 0.544 & \textbf{1.000} & 0.964 & 0.586 & 0.856 & 0.836 & 0.328 & 0.324 & 0.562 & 0.611 & 0.659 & 0.494 & 0.485 \\
orca-2 & 0.561 & 0.429 & 0.440 & 0.434 & 0.940 & 0.519 & 0.668 & 0.688 & 0.092 & 0.628 & 0.491 & 0.523 & 0.537 & 0.152 & 0.753 \\ \hline
gpt-3.5 & 0.690 & 0.360 & 0.401 & \textbf{1.000} & \textbf{1.000} & 0.389 & 0.748 & 0.744 & 0.032 & 0.032 & 0.283 & 0.486 & 0.521 & 0.062 & 0.062 \\
gpt-4 & 0.965 & \textbf{0.936} & 0.936 & \textbf{1.000} & 0.988 & 0.963 & \textbf{1.000} & \textbf{0.996} & 0.868 & \textbf{0.988} & 0.962 & \textbf{0.967} & 0.965 & 0.929 & \textbf{0.988} \\
haiku & 0.842 & 0.704 & 0.668 & 0.994 & \textbf{1.000} & 0.792 & 0.896 & 0.908 & 0.640 & 0.724 & 0.794 & 0.789 & 0.769 & 0.779 & 0.840 \\
sonnet & \textbf{0.967} & 0.929 & \textbf{0.940} & \textbf{1.000} & \textbf{1.000} & \textbf{0.965} & 0.992 & \textbf{0.996} & \textbf{0.948} & 0.924 & \textbf{0.965} & 0.959 & \textbf{0.967} & \textbf{0.973} & 0.961 \\ \bottomrule
\end{tabular}%
}
\end{table}


\subsection{Chain-of-thought prompting, one-way strategy}

Table~\ref{uni_cot} presents the results of the CoT, one-way strategy experiments.  
All models demonstrate significantly better performance compared to the analogous experiment using standard prompting. This confirms that CoT is a highly effective technique for enhancing model efficacy. 
A notable phenomenon observed here is that almost all models exhibit high precision on the same-as relationship. However, this often comes at the cost of recall, with the highest recall achieved by \model{openchat}, reaching 0.780. 


Among full open models, \model{mixtral} again outperformed the other model with an average F1-score of 0.808.
\changed{As per Section~\ref{sec:s1}, \model{mixtral}'s strong performance is likely attributed to its mixture-of-experts architecture and extensive pre-training on diverse datasets, including scientific and technical literature~\cite{jiang2024mixtral,zhao2023survey}.}

Regarding the quantised models, \model{dolphin-mistral} achieves the highest average F1-score of 0.869. It is closely followed by \model{dolphin-mistral-dpo} with a score of 0.826, \model{solar} at 0.825, and \model{openchat} at 0.823. This suggests that models fine-tuned on the OpenOrca corpus are particularly effective for this task. \changed{The OpenOrca dataset's inclusion of scientific and technical content likely contributes to this efficacy~\cite{mukherjee2023orca}.}
In this context, the use of the CoT strategy enables the quantised model \model{dolphin-mistral} to surpass the much larger \model{mixtral} model. \changed{This performance differential may stem from \model{dolphin-mistral}'s specialised fine-tuning on instruction-following datasets that emphasise reasoning processes~\cite{wang2022self}.}


\changed{Among the proprietary models, \model{gpt-4} achieved the highest performance, reaching an average F1-score of 0.911. Notably, both \model{gpt-4} and \model{sonnet} exhibit a slight decline in performance when using the CoT prompting strategy, whereas \model{haiku} and \model{gpt-3.5} show improved results under the same conditions. The counterintuitive behavior of the larger models may arise from redundancy and cognitive overhead. Since these models have already internalized complex semantic relationships during pre-training, explicitly prompting them to articulate intermediate reasoning steps could introduce unnecessary noise~\cite{wei2022chain,liu2024mind}. This additional complexity may disrupt their internal inference mechanisms, ultimately impairing performance and reducing accuracy.}

\begin{table}[!h]
\centering
\caption{Precision, recall, and F1-score for chain-of-thought prompting, one-way strategy. BR are the performance associated to \textit{broader} relationships, NA to \textit{narrower}, SA to \textit{same-as}, OT to \textit{others}, and AVG are average performance considering the four types of relationships. In \textbf{bold} are the best performing models for a given class.\label{uni_cot}}
\resizebox{\columnwidth}{!}{%
\begin{tabular}{l|rrrrr|rrrrr|rrrrr}
\toprule
\multicolumn{1}{c|}{\multirow{2}{*}{\textbf{MODEL}}} & \multicolumn{5}{c|}{\textbf{PRECISION}} & \multicolumn{5}{c|}{\textbf{RECALL}} & \multicolumn{5}{c}{\textbf{F1-SCORE}} \\
\multicolumn{1}{c|}{} & \multicolumn{1}{l}{AVG} & \multicolumn{1}{l}{BR} & \multicolumn{1}{l}{NR} & \multicolumn{1}{l}{SA} & \multicolumn{1}{l|}{OT} & \multicolumn{1}{l}{AVG} & \multicolumn{1}{l}{BR} & \multicolumn{1}{l}{NR} & \multicolumn{1}{l}{SA} & \multicolumn{1}{l|}{OT} & \multicolumn{1}{l}{AVG} & \multicolumn{1}{l}{BR} & \multicolumn{1}{l}{NR} & \multicolumn{1}{l}{SA} & \multicolumn{1}{l}{OT} \\
\midrule
mistral & 0.766 & 0.667 & 0.749 & 0.956 & 0.695 & 0.720 & 0.936 & 0.632 & 0.348 & \textbf{0.964} & 0.696 & 0.779 & 0.685 & 0.510 & 0.807 \\
mixtral & \textbf{0.849} & \textbf{0.732} & \textbf{0.807} & \textbf{1.000} & \textbf{0.858} & \textbf{0.820} & 0.948 & \textbf{0.904} & \textbf{0.488} & 0.940 & \textbf{0.808} & \textbf{0.826} & \textbf{0.853} & \textbf{0.656} & \textbf{0.897} \\
llama-2 & 0.775 & 0.466 & 0.674 & \textbf{1.000} & 0.960 & 0.612 & \textbf{0.976} & 0.860 & 0.224 & 0.388 & 0.576 & 0.630 & 0.756 & 0.366 & 0.553 \\ \hline
dolphin-mistral & \textbf{0.892} & \textbf{0.749} & 0.912 & \textbf{1.000} & 0.907 & \textbf{0.871} & \textbf{1.000} & 0.872 & 0.680 & \textbf{0.932} & \textbf{0.869} & \textbf{0.856} & \textbf{0.892} & 0.810 & \textbf{0.919} \\
dolphin-mistral-dpo & 0.863 & 0.685 & \textbf{0.947} & 0.994 & 0.827 & 0.827 & 0.992 & 0.708 & 0.688 & 0.920 & 0.826 & 0.811 & 0.810 & 0.813 & 0.871 \\
dolphin-openorca & 0.840 & 0.524 & 0.865 & \textbf{1.000} & 0.971 & 0.736 & \textbf{1.000} & 0.848 & 0.560 & 0.536 & 0.738 & 0.688 & 0.857 & 0.718 & 0.691 \\
openchat & 0.844 & 0.692 & 0.874 & 0.947 & 0.864 & 0.822 & 0.988 & 0.780 & \textbf{0.780} & 0.740 & 0.823 & 0.814 & 0.825 & \textbf{0.855} & 0.797 \\
openchat-gemma & 0.822 & 0.429 & 0.864 & \textbf{1.000} & \textbf{0.993} & 0.650 & \textbf{1.000} & 0.408 & 0.588 & 0.604 & 0.662 & 0.600 & 0.554 & 0.741 & 0.751 \\
solar & 0.858 & 0.739 & 0.771 & \textbf{1.000} & 0.924 & 0.830 & 0.940 & \textbf{0.928} & 0.580 & 0.872 & 0.825 & 0.828 & 0.842 & 0.734 & 0.897 \\
mistral-openorca & 0.810 & 0.551 & 0.865 & 0.987 & 0.836 & 0.736 & \textbf{1.000} & 0.744 & 0.608 & 0.592 & 0.739 & 0.710 & 0.800 & 0.753 & 0.693 \\
eurus & 0.796 & 0.661 & 0.635 & \textbf{1.000} & 0.887 & 0.723 & \textbf{1.000} & 0.912 & 0.416 & 0.564 & 0.706 & 0.796 & 0.749 & 0.588 & 0.690 \\
llama-3 & 0.723 & 0.547 & 0.491 & \textbf{1.000} & 0.852 & 0.626 & 0.696 & 0.748 & 0.232 & 0.828 & 0.605 & 0.613 & 0.593 & 0.377 & 0.840 \\
orca-2 & 0.787 & 0.622 & 0.624 & \textbf{1.000} & 0.904 & 0.711 & 0.868 & \textbf{0.928} & 0.404 & 0.644 & 0.700 & 0.725 & 0.746 & 0.576 & 0.752 \\ \hline
gpt-3.5 & 0.727 & 0.534 & 0.501 & 0.896 & 0.976 & 0.573 & 0.904 & 0.880 & 0.344 & 0.164 & 0.522 & 0.672 & 0.639 & 0.497 & 0.281 \\
gpt-4 & \textbf{0.924} & \textbf{0.828} & \textbf{0.921} & \textbf{1.000} & 0.945 & \textbf{0.914} & 0.984 & \textbf{0.980} & \textbf{0.724} & \textbf{0.968} & \textbf{0.911} & \textbf{0.899} & \textbf{0.950} & \textbf{0.840} & \textbf{0.957} \\
haiku & 0.861 & 0.664 & 0.809 & \textbf{1.000} & \textbf{0.969} & 0.812 & \textbf{0.996} & 0.952 & 0.540 & 0.760 & 0.806 & 0.797 & 0.875 & 0.701 & 0.852 \\
sonnet & 0.909 & \textbf{0.821} & 0.865 & \textbf{1.000} & 0.950 & 0.896 & \textbf{0.992} & 0.976 & 0.696 & 0.920 & 0.893 & \textbf{0.899} & 0.917 & 0.821 & 0.935 \\ \bottomrule
\end{tabular}%
}
\end{table}

\subsection{Chain-of-thought prompting, two-way strategy}

The values are generally consistent with those obtained from the CoT one-way strategy experiments (see Table~\ref{uni_cot}). Specifically, \model{mixtral}, \model{dolphin-mistral}, and \model{gtp-4} emerge as the top-performing models among the full open models, the quantised models, and the proprietary models, respectively. 
The average F1 score increases by a few percentage points in 15 out of 17 models, confirming that the two-way strategy remains effective when combined with CoT. 
\changed{
The consistent improvement across models when employing the two-way strategy aligns with previous findings demonstrating that techniques aimed at enhancing logical consistency can lead to more robust judgments \cite{liu2024aligning}. }



\changed{For quantised models, the good performance of \model{dolphin-mistral} further suggests that pretraining on OpenOrca~\cite{mukherjee2023orca} yields a consistent advantage on this task. In this case, the quantisation process~\cite{dettmers2022gpt3} appears to preserve the model’s core representational capacity for handling technical terminology. This preservation of performance post-quantisation is particularly noteworthy and indicates that the model's parameters encode knowledge efficiently, even under reduced precision constraints~\cite{frantar2023optq}.
}

\changed{The variation in performance across different versions of the same base model (e.g., different \model{mistral} derivatives) suggests that fine-tuning procedures can either enhance or diminish domain-specific capabilities present in the base model~\cite{chang2024survey}. This highlights the trade-offs inherent in optimising for general capabilities versus domain-specific knowledge retention~\cite{zhou2023lima}. The consistent improvement achieved through two-way verification with CoT reasoning across nearly all models suggests that, regardless of pre-training, explicit reasoning helps models more effectively leverage their knowledge of technical terminology and scientific concepts~\cite{zheng2023judging, miao2023selfcheck}.}

\begin{table}[!h]
\centering
\caption{Precision, recall, and F1-score for chain-of-thought prompting, two-way strategy. BR are the performance associated to \textit{broader} relationships, NA to \textit{narrower}, SA to \textit{same-as}, OT to \textit{others}, and AVG are average performance considering the four types of relationships. In \textbf{bold} are the best performing models for a given class.\label{bid_cot}}
\resizebox{\columnwidth}{!}{%
\begin{tabular}{l|rrrrr|rrrrr|rrrrr}
\toprule
\multicolumn{1}{c|}{\multirow{2}{*}{\textbf{MODEL}}} & \multicolumn{5}{c|}{\textbf{PRECISION}} & \multicolumn{5}{c|}{\textbf{RECALL}} & \multicolumn{5}{c}{\textbf{F1-SCORE}} \\
\multicolumn{1}{c|}{} & \multicolumn{1}{l}{AVG} & \multicolumn{1}{l}{BR} & \multicolumn{1}{l}{NR} & \multicolumn{1}{l}{SA} & \multicolumn{1}{l|}{OT} & \multicolumn{1}{l}{AVG} & \multicolumn{1}{l}{BR} & \multicolumn{1}{l}{NR} & \multicolumn{1}{l}{SA} & \multicolumn{1}{l|}{OT} & \multicolumn{1}{l}{AVG} & \multicolumn{1}{l}{BR} & \multicolumn{1}{l}{NR} & \multicolumn{1}{l}{SA} & \multicolumn{1}{l}{OT} \\
\midrule
mistral & 0.830 & \textbf{0.775} & 0.734 & 0.932 & 0.880 & 0.810 & 0.936 & 0.904 & 0.492 & \textbf{0.908} & 0.799 & 0.848 & 0.810 & 0.644 & 0.894 \\
mixtral & \textbf{0.876} & 0.768 & \textbf{0.795} & \textbf{1.000} & 0.940 & \textbf{0.850} & \textbf{0.992} & \textbf{0.948} & \textbf{0.576} & 0.884 & \textbf{0.843} & \textbf{0.866} & \textbf{0.865} & \textbf{0.731} & \textbf{0.911} \\
llama-2 & 0.766 & 0.502 & 0.563 & \textbf{1.000} & \textbf{1.000} & 0.588 & 0.936 & 0.928 & 0.280 & 0.208 & 0.534 & 0.654 & 0.701 & 0.438 & 0.344 \\ \hline
dolphin-mistral & \textbf{0.930} & \textbf{0.874} & \textbf{0.852} & \textbf{1.000} & 0.995 & \textbf{0.920} & \textbf{0.996} & \textbf{0.992} & 0.844 & 0.848 & \textbf{0.920} & \textbf{0.931} & \textbf{0.917} & \textbf{0.915} & 0.916 \\
dolphin-mistral-dpo & 0.915 & 0.859 & 0.831 & 0.986 & 0.982 & 0.905 & 0.952 & 0.964 & 0.840 & \textbf{0.864} & 0.906 & 0.903 & 0.893 & 0.907 & \textbf{0.919} \\
dolphin-openorca & 0.828 & 0.640 & 0.670 & \textbf{1.000} & \textbf{1.000} & 0.745 & 0.968 & 0.968 & 0.704 & 0.340 & 0.724 & 0.771 & 0.792 & 0.826 & 0.508 \\
openchat & 0.857 & 0.763 & 0.777 & 0.903 & 0.986 & 0.834 & 0.940 & 0.932 & \textbf{0.892} & 0.572 & 0.828 & 0.842 & 0.847 & 0.897 & 0.724 \\
openchat-gemma & 0.798 & 0.601 & 0.597 & 0.994 & \textbf{1.000} & 0.711 & 0.868 & 0.852 & 0.672 & 0.452 & 0.709 & 0.710 & 0.702 & 0.802 & 0.623 \\
solar & 0.886 & 0.759 & 0.815 & \textbf{1.000} & 0.971 & 0.861 & 0.980 & 0.968 & 0.704 & 0.792 & 0.860 & 0.855 & 0.885 & 0.826 & 0.872 \\
mistral-openorca & 0.846 & 0.686 & 0.698 & \textbf{1.000} & \textbf{1.000} & 0.784 & 0.968 & 0.972 & 0.756 & 0.440 & 0.772 & 0.803 & 0.813 & 0.861 & 0.611 \\
eurus & 0.833 & 0.672 & 0.670 & \textbf{1.000} & 0.990 & 0.756 & 0.992 & \textbf{0.992} & 0.628 & 0.412 & 0.739 & 0.801 & 0.800 & 0.772 & 0.582 \\
llama-3 & 0.766 & 0.556 & 0.541 & \textbf{1.000} & 0.968 & 0.661 & 0.808 & 0.812 & 0.292 & 0.732 & 0.649 & 0.659 & 0.650 & 0.452 & 0.834 \\
orca-2 & 0.809 & 0.598 & 0.651 & \textbf{1.000} & 0.985 & 0.724 & 0.904 & 0.912 & 0.544 & 0.536 & 0.720 & 0.720 & 0.760 & 0.705 & 0.694 \\ \hline
gpt-3.5 & 0.741 & 0.547 & 0.553 & 0.863 & \textbf{1.000} & 0.606 & 0.904 & 0.920 & 0.528 & 0.072 & 0.540 & 0.682 & 0.691 & 0.655 & 0.134 \\
gpt-4 & \textbf{0.941} & \textbf{0.899} & \textbf{0.889} & \textbf{1.000} & 0.975 & \textbf{0.935} & \textbf{1.000} & \textbf{0.996} & \textbf{0.792} & \textbf{0.952} & \textbf{0.933} & \textbf{0.947} & \textbf{0.940} & \textbf{0.884} & \textbf{0.964} \\
haiku & 0.877 & 0.755 & 0.765 & \textbf{1.000} & 0.989 & 0.842 & 0.984 & 0.988 & 0.680 & 0.716 & 0.839 & 0.854 & 0.862 & 0.809 & 0.831 \\
sonnet & 0.916 & 0.822 & 0.855 & \textbf{1.000} & 0.986 & 0.901 & \textbf{1.000} & 0.992 & 0.768 & 0.844 & 0.900 & 0.902 & 0.918 & 0.869 & 0.909 \\ \bottomrule
\end{tabular}%
}
\end{table}


\subsection{Comparative Analysis}

Table~\ref{tab:overall-f1-score} presents the average F1-scores for all 17 models across the four strategies, including the relative differences between one-way and two-way strategies for both standard and CoT prompting.
The results indicate that shifting from a one-way to a two-way strategy improved performance in 14 models for standard prompting and in 15 models for CoT prompting.

Generally, we observe a gradual increase in the average F1-score from left to right in Table~\ref{tab:overall-f1-score}, demonstrating the beneficial effects of both CoT and the two-way strategy. 
Thirteen models showed overall improvement, while four models (\model{llama-2}, \model{dolphin-openorca}, \model{gpt-4}, and \model{sonnet}) experienced a decline in performance.
When applying a 0.7 average F1-score as a quality benchmark, only \model{dolphin-openorca} and \model{openchat} surpassed this threshold under a standard prompting approach with a one-way strategy. However, with a two-way strategy in standard prompting, five models exceeded the threshold. This number increased to eight models when a CoT, one-way strategy was used. Ultimately, nine models achieved an average F1-score above the 0.7 threshold when employing a CoT, two-way strategy. 
These results underscore the critical importance of a good prompting strategy when working in a zero-shot setting.



Finally, we can analyse the best model for each of the three categories: fully open models, quantised models, and proprietary models. The \model{mixtral} model stands out among fully open models, particularly when used with the standard prompting and two-way strategy (0.847 F1). In contrast, the \model{dolphin-mistral} model surpasses other quantised models when applied with the CoT prompting and two-way strategy (0.920 F1).  Notably, \model{dolphin-mistral} outperforms \model{mixtral}, likely due to the additional fine-tuning on datasets like OpenOrca, which seem particularly beneficial for this task.
Among proprietary models, \model{sonnet} delivers the best performance when using standard prompts and a straightforward, one-way approach (0.967 F1). Interestingly, unlike most open models, its performance declines with more complex prompts. Since \model{sonnet} is a black-box system that may incorporate additional components, pinpointing the exact cause of this behaviour is challenging.

\begin{table}[!h]
\centering
\caption{Average F1-scores obtained from the 17 models on all experiments.  In \textbf{bold} are the best performing models for a given strategy.\label{tab:overall-f1-score}}
\resizebox{0.8\columnwidth}{!}{%
\begin{tabular}{l|r|r|r|r|r|r|r}
\toprule
\multicolumn{1}{c|}{\multirow{2}{*}{\textbf{MODEL}}} & \multicolumn{3}{c|}{\textbf{Standard Prompting}} & \multicolumn{3}{c}{\textbf{CoT Prompting}} & \multicolumn{1}{|l}{\textbf{Overall}} \\ 
& \multicolumn{1}{l|}{\textit{One-way}} & \multicolumn{1}{l|}{\textit{Two-way}} & \multicolumn{1}{l|}{\textit{Difference}} & \multicolumn{1}{l|}{\textit{One-way}} & \multicolumn{1}{l|}{\textit{Two-way}} & \multicolumn{1}{l}{\textit{Difference}}  & \multicolumn{1}{|l}{\textit{Improvement}}     \\ \midrule
mistral                                             & 0.313                                & 0.318                                & 0.005                                   & 0.696                                & 0.799                                & \textbf{0.103}                          & 0.486      \\
mixtral                                             & \textbf{0.779}                       & \textbf{0.847}                       & \textbf{0.068}                          & \textbf{0.808}                       & \textbf{0.843}                       & 0.035                                   & 0.068      \\
llama-2                                             & 0.669                                & 0.605                                & -0.064                                  & 0.576                                & 0.534                                & -0.042                                  & 0.135      \\
\midrule
dolphin-mistral                                     & 0.599                                & 0.706                                & 0.107                                   & \textbf{0.869}                       & \textbf{0.920}                       & 0.051                                   & 0.321      \\
dolphin-mistral-dpo                                 & 0.552                                & 0.795                                & \textbf{0.243}                          & 0.826                                & 0.906                                & \textbf{0.080}                          & 0.354      \\
dolphin-openorca                                    & \textbf{0.724}                       & \textbf{0.853}                       & 0.129                                   & 0.738                                & 0.724                                & -0.014                                  & 0.129      \\
openchat                                            & 0.705                                & 0.783                                & 0.078                                   & 0.823                                & 0.828                                & 0.005                                   & 0.123      \\
openchat-gemma                                      & 0.506                                & 0.579                                & 0.073                                   & 0.662                                & 0.709                                & 0.047                                   & 0.203      \\
solar                                               & 0.646                                & 0.710                                & 0.064                                   & 0.825                                & 0.860                                & 0.035                                   & 0.214      \\
mistral-openorca                                    & 0.613                                & 0.630                                & 0.017                                   & 0.739                                & 0.772                                & 0.033                                   & 0.159      \\
eurus                                               & 0.466                                & 0.488                                & 0.022                                   & 0.706                                & 0.739                                & 0.033                                   & 0.273      \\
llama-3                                             & 0.568                                & 0.562                                & -0.006                                  & 0.605                                & 0.649                                & 0.044                                   & 0.087      \\
orca-2                                              & 0.372                                & 0.491                                & 0.119                                   & 0.700                                & 0.720                                & 0.020                                   & 0.348      \\
\midrule
gpt-3.5                                             & 0.195                                & 0.283                                & \textbf{0.088}                          & 0.522                                & 0.540                                & 0.018                                   & 0.345      \\
gpt-4                                               & 0.948                                & 0.962                                & 0.014                                   & \textbf{0.911}                       & \textbf{0.933}                       & 0.022                                   & 0.051      \\
haiku                                               & 0.712                                & 0.794                                & 0.082                                   & 0.806                                & 0.839                                & \textbf{0.033}                          & 0.127      \\
sonnet                                              & \textbf{0.967}                       & \textbf{0.965}                       & -0.002                                  & 0.893                                & 0.900                                & 0.007                                   & 0.074     \\
\bottomrule
\end{tabular}
}
\end{table}

\section{Error Analysis}\label{erroranalysis}

\begin{figure*}[t!]
    \centering
    \begin{subfigure}[t]{0.50\textwidth}
        \centering
        \includegraphics[width=.95\linewidth]{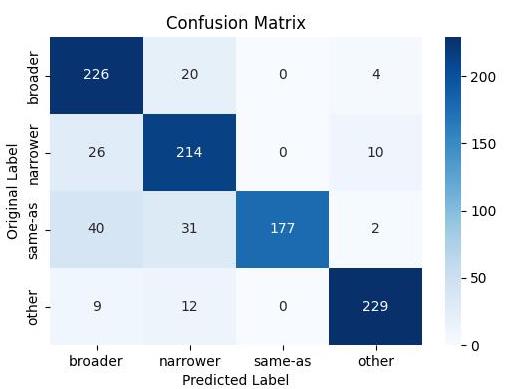}
        \caption{\model{mixtral} in standard prompting, \\two-way strategy.}
    \end{subfigure}%
    \begin{subfigure}[t]{0.50\textwidth}
        \centering
        \includegraphics[width=.95\linewidth]{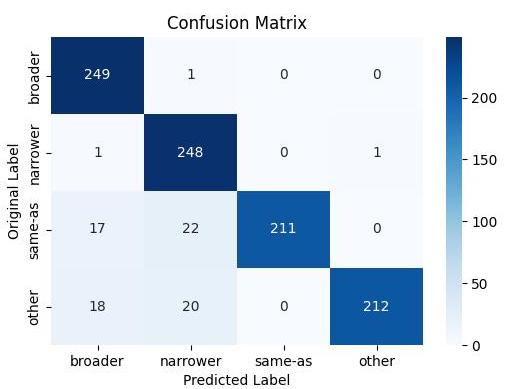}
        \caption{\model{dolphin-mistral} in CoT prompting, \\two-way strategy.}
    \end{subfigure}
    \begin{subfigure}[t]{0.50\textwidth}
        \centering
        \includegraphics[width=.95\linewidth]{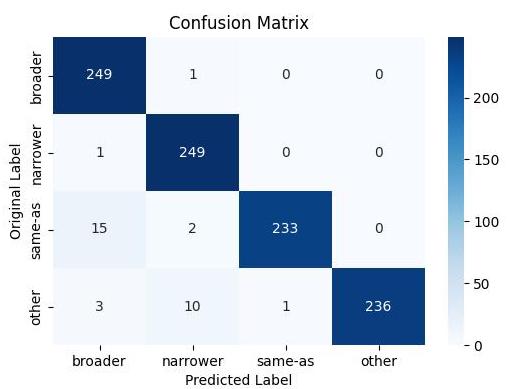}
        \caption{\model{sonnet} in standard prompting, \\one-way strategy.}
    \end{subfigure}
    \caption{Confusion matrices: a) for the best open model, b) for the best quantised model, and c) for the best proprietary model.}
    \label{fig:confusionmatrices}
\end{figure*}

In this section, we examine common error types displayed by the top-performing LLMs across the three categories: a) \model{mixtral} (standard prompting, two-way), b) \model{dolphin-mistral} (CoT, two-way), and c) \model{sonnet} (standard prompting, one-way). Figure~\ref{fig:confusionmatrices} reports the relevant confusion matrices.



The \model{mixtral} model misclassified 154 out of 1,000 topic pairs. Specifically, 75 pairs were incorrectly identified as broader, 63 as narrower, 16 as other, and none as same-as. \model{mixtral} demonstrates a highly conservative approach in predicting the same-as category, achieving a precision of 1, as indicated in Table~\ref{bid_sta}. This comes at a cost, as the model exhibits a tendency to misclassify pairs that are labelled as same-as in the gold standard, often categorising them incorrectly as either broader or narrower. This misclassification yields a low recall for the same-as category (0.708), which represents a significant limitation of the model. Specifically, 29.2\% of same-as pairs were misclassified: 16\% were labelled as broader, 12.4\% as narrower, and 0.8\% as other.
Additionally, the \model{mixtral} model occasionally confuses broader pairs with narrower ones and vice versa. Notably, 9.6\% of broader pairs were incorrectly predicted, with 8\% labelled as narrower and 1.6\% as other. Similarly, 14.4\% of narrower pairs were misclassified, with 10.4\% labelled as broader and 4\% as other.







The \model{dolphin-mistral} model exhibits a different pattern compared to \model{mixtral}, particularly excelling in pairs labelled as broader and narrower in the gold standard. It makes 80 misclassifications, predominantly in the same-as (39) and other (38) categories. Although the model still faces challenges with same-as relationships, it performs better than \model{mixtral}. Specifically, for the same-as relationships, a small percentage (6.8\%) are misclassified as broader, while 8.8\% are incorrectly identified as narrower.
Notably, the other relationships are misclassified equally as broader or narrower.







The \model{sonnet} model exhibits a similar performance pattern to \model{dolphin-mistral}, achieving near-perfect accuracy across both broader and narrow categories, while showing notable improvements in the other two categories. The model made only 33 classification errors, with the majority occurring in same-as (17) and other (14). Specifically, within the same-as misclassifications, 6\% were incorrectly identified as broader, and 0.8\% as narrower. In contrast, for the other category, 1.6\% of errors were misclassified as broader or same-as, while 4\% were erroneously categorised as other.


The error analysis provides several valuable insights. First, the best-performing model excels in identifying hierarchical relationships, such as broader and narrower.  This is a promising indication of the potential for LLMs to assist in ontology generation, where hierarchical relationships are fundamental to structuring the ontology.
Second, while the very best model shows reasonable success in identifying ``same-as'' relationships, this task proves challenging. This difficulty may arise because same-as relationships between research topics are not purely syntactic synonyms and need deep expertise in the field to be interpreted correctly. Fine-tuning the model on task-specific data could help address this issue, and we plan to explore this approach in future work.
 \color{black}

\changedsec{
\section{Implications and limitations}\label{sec:limitations}
}

\changedsec{

The experiments presented in this study led to several notable insights. 
\textit{First}, many models demonstrate strong zero-shot reasoning capabilities for this task, at least within the domain of engineering and computer science as represented by the IEEE Thesaurus. This performance may be partly due to the nature of relevant concepts in electrical engineering and computer science, which are likely well represented in the training data of LLMs, including code repositories, technical documentation, manuals, and patents. We intend to extend our analysis to other disciplines to determine whether this strong performance is consistent across fields and to evaluate the potential benefits of fine-tuning models for this specific task.
\textit{Second}, the performance of LLMs is strongly influenced by the choice of prompting strategy. In particular, CoT prompting clearly outperforms standard prompting, aligning with findings from prior research~\cite{wei2022chain}. 
Additionally, running the classification process twice, switching the topic position, and applying simple rules to integrate the two outputs further reduces the number of errors.
Indeed, the two-way CoT variant outperforms all other approaches, especially for smaller, more scalable, and cost-efficient models. The analysis indicates that selecting appropriate prompting strategies can lead to notable performance improvements, with gains exceeding 0.2 F1 points in several cases (see Table~\ref{tab:overall-f1-score}, last column).
\textit{Third}, smaller models achieve performance comparable to that of the best-performing architectures when adopting the best prompting strategy. Such models, particularly when fine-tuned on relevant datasets, can be highly effective for this task and may even outperform their base versions. For example, \model{dolphin-mistral} and \model{dolphin-mistral-dpo} surpass the standard \model{mistral} and are only outperformed by the largest proprietary models. These results suggest promising directions for the development of optimised and efficient models for ontology generation.

}


\changedsec{
While we believe that these insights represent a substantial advancement in the state of the art, our current research settings present certain limitations that warrant discussion. We plan to address and explore these further in future work.

The first limitation lies in the domain-specific nature of our findings, which, as previously discussed, are currently confined to the IEEE taxonomy and the engineering field. A broader evaluation is required to assess the performance of LLMs in other domains. To this end, we are actively working on expanding our dataset to include classifications from additional disciplines, such as the Medical Subject Headings and the Physics Subject Headings, which represent the domains of biomedicine and physics, respectively. 
A second limitation is that our evaluation only focuses on LLM-based strategies. We do not compare these approaches with traditional AI methods, such as rule-based, statistical, or hybrid techniques. Nevertheless, we recognise the importance of such comparisons and intend to include them in future experiments.
A third limitation arises from the dynamic nature of scientific research, where disciplines are continuously redefined by emerging paradigms and discoveries. Although our current results provide valuable insights, LLMs trained on present-day data may struggle to accurately classify relationships involving topics that have not yet emerged. Therefore, while our methodological approach offers long-term relevance, the specific LLMs employed will need to be updated periodically to reflect the most recent developments. We posit that future generations of LLMs will likely continue the observed trend of improvement, offering greater efficiency and enhanced performance at a reduced computational cost.
}

In addition, while this paper focused exclusively on identifying relationships among a subset of topics, future work will need to address the construction of a complete ontology. This entails resolving several challenges, including the evaluation of relationships across a much larger set of topics. To improve scalability and efficiency, it is crucial to identify strategies that reduce the search space. One promising approach proposed in the literature~\cite{pisu2024leveraging,pisu2024classifying} involves leveraging topic co-occurrence patterns in research paper abstracts and titles. Topic pairs that rarely or never co-occur can be automatically classified under the ``other'' category, thereby significantly reducing the computational cost of model inference.

Finally, while the results produced by LLMs demonstrate high accuracy, in realistic settings they must be further verified and refined by human experts.  We envision a collaborative workflow in which LLMs generate initial ontology drafts that are subsequently reviewed and refined by domain experts. This process ensures both accuracy and the preservation of critical human judgment, which remains essential for developing robust and reliable KOS. 
Recent studies have proposed various strategies for integrating human-in-the-loop methodologies into the construction of knowledge graphs and ontologies~\cite{tsaneva2025knowledge}.

\vspace{0.5cm}

\section{Conclusions}\label{sec:conclusions}
This paper conducted an in-depth analysis of the capability of LLMs to identify semantic relationships between research topics for the purpose of ontology generation. We evaluated seventeen LLMs, categorised into three groups: fully open models (3), open quantised models (10), and proprietary models (4). We evaluated their performance against a gold standard of 1,000 relationships from the IEEE Thesaurus. We also investigated four distinct zero-shot prompting strategies, which yielded significantly different outcomes.

Our results demonstrated the impressive zero-shot reasoning capabilities of state-of-the-art models. Three models emerged as top performers within their respective categories: Mixtral-8×7B (\model{mixtral}: 0.847 average F1-score) for the full open models; Dolphin-2.1-Mistral-7B (\model{dolphin-mistral}: 0.920) for open quantised models; and Claude 3 Sonnet (\model{sonnet}: 0.967) for proprietary models. 

We are currently working on several fronts. We are fine-tuning off-the-shelf models like BERT, Mistral, LLaMa 3, Flan T5, Gemma, Phi 3, and DeepSeek with the aim of enhancing the performance of lightweight models to reduce computational costs and environmental impact when generating ontologies of research areas. \changed{
Furthermore, we plan to incorporate a semantic reasoner, such as RDFox~\cite{nenov2015rdfox}, into our pipeline to identify and resolve inconsistencies. We will also make all the resulting ontologies publicly available, formatted according to the SKOS standard. Other directions include exploring explainability techniques, including SHAP values~\cite{marcilio2020explanations}, attention visualization~\cite{vig-2019-multiscale}, and the identification of monosemantic features~\cite{templeton2024scaling} to understand how LLMs process ontological relationships. } 
\changed{We also plan to investigate potential biases in LLM-generated ontologies, such as examining whether inference error rates vary across different sub-domains of Engineering. Additionally, we will benchmark our findings against additional state-of-the-art ontology generation methods, including rule-based systems and statistical approaches.
Ultimately, our goal is to develop a comprehensive, cross-disciplinary ontology of research topics. To this end, we plan to expand our analysis to cover a variety of scientific domains, including materials science, medicine, and physics.}

\textbf{Declaration of generative AI and AI-assisted technologies in the writing process}

During the preparation of this work the author(s) used Grammarly in order to improve readability. After using this tool/service, the author(s) reviewed and edited the content as needed and take(s) full responsibility for the content of the publication.

\section*{Acknowledgements}
This publication is based upon work from COST Action CA23147 GOBLIN - Global Network on Large-Scale, Cross-domain and Multilingual Open Knowledge Graphs, supported by COST (European Cooperation in Science and Technology, \url{https://www.cost.eu}).

\bibliographystyle{elsarticle-num} 
\bibliography{bib}

\newpage

\appendix

\section{Prompt for Standard Prompting}\label{app:promtpstandard}
For consistency we applied the same prompt across all the models. We engineered this prompt through various refinements, to ensure optimal comprehension of the task and accurate responses.
Below is the template of our prompt, customised for each topic pair by substituting [TOPIC-A] for the first topic and [TOPIC-B] for the second.

\begin{Verbatim}[breaklines=true]
Classify the relationship between '[TOPIC-A]' and '[TOPIC-B]' by applying the following relationship definitions:
1. '[TOPIC-A]' is-broader-than '[TOPIC-B]' if '[TOPIC-A]' is a super-category of '[TOPIC-B]', that is '[TOPIC-B]' is a type, a branch, or a specialised aspect of '[TOPIC-A]' or that '[TOPIC-B]' is a tool or a methodology mostly used in the context of '[TOPIC-A]' (e.g., car is-broader-than wheel).
2. '[TOPIC-A]' is-narrower-than '[TOPIC-B]' if '[TOPIC-A]' is a sub-category of '[TOPIC-B]', that is '[TOPIC-A]' is a type, a branch, or a specialised aspect of '[TOPIC-B]' or that '[TOPIC-A]' is a tool or a methodology mostly used in the context of '[TOPIC-B]' (e.g., wheel is-narrower-than car).
3. '[TOPIC-A]' is-same-as-than '[TOPIC-B]' if '[TOPIC-A]' and '[TOPIC-B]' are synonymous terms denoting an identical concept (e.g., beautiful is-same-as-than attractive), including when one is the plural form of the other (e.g., cat is-same-as-than cats).
4. '[TOPIC-A]' is-other-than '[TOPIC-B]' if '[TOPIC-A]' and '[TOPIC-B]' either have no direct relationship or share a different kind of relationship that does not fit into the other defined relationships.

Given the previous definitions, determine which one of the following statements is correct:
1. '[TOPIC-A]' is-broader-than '[TOPIC-B]'
2. '[TOPIC-B]' is-narrower-than '[TOPIC-A]'
3. '[TOPIC-A]' is-narrower-than '[TOPIC-B]'
4. '[TOPIC-B]' is-broader-than '[TOPIC-A]'
5. '[TOPIC-A]' is-same-as-than '[TOPIC-B]'
6. '[TOPIC-A]' is-other-than '[TOPIC-B]'

Answer by only stating the correct statement and its number.
\end{Verbatim}

\section{Prompt for Chain-of-Thought Prompting}\label{app:promtpcot}
In the CoT prompting we have a two-phase interaction with the model. In the first interaction the model is sked to provide a definition of both topics, formulate a sentence that incorporates both topics, and finally discuss their potential semantic relationships. The resulting response is then fed back into the model in the second phase, with the sole task of determining the type of semantic relationship. Below are our prompt templates, customised for each topic pair by substituting [TOPIC-A] for the first topic and [TOPIC-B] for the second.

\textbf{First interaction}
\begin{Verbatim}[breaklines=true]
Classify the relationship between '[TOPIC-A]' and '[TOPIC-B]' by applying the following relationship definitions:
1. '[TOPIC-A]' is-broader-than '[TOPIC-B]' if '[TOPIC-A]' is a super-category of '[TOPIC-B]', that is '[TOPIC-B]' is a type, a branch, or a specialised aspect of '[TOPIC-A]' or that '[TOPIC-B]' is a tool or a methodology mostly used in the context of '[TOPIC-A]' (e.g., car is-broader-than wheel).
2. '[TOPIC-A]' is-narrower-than '[TOPIC-B]' if '[TOPIC-A]' is a sub-category of '[TOPIC-B]', that is '[TOPIC-A]' is a type, a branch, or a specialised aspect of '[TOPIC-B]' or that '[TOPIC-A]' is a tool or a methodology mostly used in the context of '[TOPIC-B]' (e.g., wheel is-narrower-than car).
3. '[TOPIC-A]' is-same-as-than '[TOPIC-B]' if '[TOPIC-A]' and '[TOPIC-B]' are synonymous terms denoting a very similar concept (e.g., 'beautiful' is-same-as-than 'attractive'), including when one is the plural form of the other (e.g., cat is-same-as-than cats).
4. '[TOPIC-A]' is-other-than '[TOPIC-B]' if '[TOPIC-A]' and '[TOPIC-B]' either have no direct relationship or share a different kind of relationship that does not fit into the other defined relationships.

Think step by step by following these sequential instructions:
1) Provide a precise definition for '[TOPIC-A]'.
2) Provide a precise definition for '[TOPIC-B]'.
3) Formulate a sentence that includes both '[TOPIC-A]' and '[TOPIC-B]'.
4) Discuss '[TOPIC-A]' and '[TOPIC-B]' usage and relationship (is-narrower-than, is-broader-than, is-same-as-than, or is-other-than).
\end{Verbatim}

\bigskip

\textbf{Second interaction}
\begin{Verbatim}[breaklines=true]
[PREVIOUS-RESPONSE]
Given the previous discussion, determine which one of the following statements is correct:
1. '[TOPIC-A]' is-broader-than '[TOPIC-B]'
2. '[TOPIC-B]' is-narrower-than '[TOPIC-A]'
3. '[TOPIC-A]' is-narrower-than '[TOPIC-B]'
4. '[TOPIC-B]' is-broader-than '[TOPIC-A]'
5. '[TOPIC-A]' is-same-as-than '[TOPIC-B]'
6. '[TOPIC-A]' is-other-than '[TOPIC-B]'

Answer by only stating the number of the correct statement.
\end{Verbatim}



\end{document}